\documentclass[acmsmall,screen]{acmart}
\usepackage{subfloat}
\usepackage{subcaption}
\usepackage{xcolor}
\newcommand{\markit}[1]{\textcolor{blue}{#1}}

\AtBeginDocument{%
  \providecommand\BibTeX{{%
    \normalfont B\kern-0.5em{\scshape i\kern-0.25em b}\kern-0.8em\TeX}}}

\begin{document}

\title{A multidimensional measurement of photorealistic avatar quality of experience}


\author{Ross Cutler}
\author{Babak Naderi}
\author{Vishak Gopal}
\author{Dharmendar Palle}


\begin{abstract}
Photorealistic avatars are human avatars that look, move, and talk like real people. The performance of photorealistic avatars has significantly improved recently based on objective metrics such as PSNR, SSIM,  LPIPS, FID, and FVD. However, recent photorealistic avatar publications do not provide subjective tests of the avatars to measure human usability factors. We provide an open source test framework to subjectively measure photorealistic avatar performance in ten dimensions: realism, trust, comfortableness using, comfortableness interacting with, appropriateness for work, creepiness, formality, affinity, resemblance to the person, and emotion accuracy. Using telecommunication scenarios, we show that the correlation of nine of these subjective metrics with PSNR, SSIM, LPIPS, FID, and FVD is weak, and moderate for emotion accuracy. The crowdsourced subjective test framework is highly reproducible and accurate when compared to a panel of experts. We analyze a wide range of avatars from photorealistic to cartoon-like and show that some photorealistic avatars are approaching real video performance based on these dimensions. We also find that for avatars above a certain level of realism, eight of these measured dimensions are strongly correlated. This means that avatars that are not as realistic as real video will have lower trust, comfortableness using, comfortableness interacting with, appropriateness for work, formality, and affinity, and higher creepiness compared to real video. In addition, because there is a strong linear relationship between avatar affinity and realism, there is no uncanny valley effect for photorealistic avatars in the telecommunication scenario. We suggest several extensions of this test framework for future work and discuss design implications for telecommunication systems. The test framework is available at \url{https://github.com/microsoft/P.910}.
\end{abstract}





\maketitle

\section{Introduction}
\label{sec:introduction}
Photorealistic avatars are human avatars that look, move, and talk like real people. Photorealistic avatars can be used for various applications, such as: 

\begin{itemize}
    \item Telecommunication: Photorealistic avatars can be used instead of two-dimensional webcam videos to create a virtual meeting space. Users can interact while maintaining correct eye gaze, allowing participants to know who is looking at whom, which increases trust \cite{nguyen_multiview_2007} while reducing video fatigue by reducing hyper-gaze \cite{gale_eeg_1975,fauville_nonverbal_2021}. 
    \item Health care: Photorealistic avatars can be used to provide virtual consultations, training, therapy, and education for patients and medical professionals. For example, a photorealistic avatar of a doctor can explain a diagnosis, prescribe a treatment, or demonstrate a procedure to a patient.
    \item Education: Photorealistic avatars can be used to create immersive, personalized, and interactive learning environments for students and teachers. For example, a photorealistic avatar of a teacher can guide students through a lesson, provide feedback, or answer questions.
    \item Retail and e-commerce: Photorealistic avatars can be used to enhance the (online) shopping experience for customers, sellers, and customer service. For customers, photorealistic avatars can enable virtual try-on experiences for clothing, accessories, or makeup, allowing customers to visualize how products will look on them before making a purchase.
    \item Entertainment: Photorealistic avatars can be used to create realistic and engaging characters for games, movies, shows, and social media. For example, a photorealistic avatar of an actor can perform scenes, interact with fans, or promote brands.
\end{itemize}

While there are many applications of photorealistic avatars, our focus is on evaluating them for the telecommunication scenario to improve trust and reduce video fatigue. Therefore the avatar test sequences we use are targeted for telecommunication, which includes people talking and expressing a wide range of emotions (happy, sad, surprised, fear, anger, and disgust). However, we think that the test framework and methodology is applicable to all of the above mentioned avatar applications. 

The performance of photorealistic avatars has been improving significantly recently based on objective metrics such as PSNR \cite{gonzalez_digital_2006}, SSIM \cite{wang_image_2004}, LPIPS \cite{zhang_unreasonable_2018}, FID \cite{heusel_gans_2017}, and FVD \cite{unterthiner_fvd_2019}. However, recent avatar publications do not provide subjective tests of the avatars to measure human usability factors (e.g., \cite{saito_relightable_2024,tian_emo_2024,xu_gaussian_2024,huang_make-your-anchor_2024,shao_splattingavatar_2024,zhou_ultravatar_2024,deng_portrait4d_2024,kirschstein_diffusionavatars_2024,xu_vasa-1_2024,liu_disentangling_2024}). We believe this is because such subjective tests are challenging, but also because there is no standardized or readily available method to do so. The usability factors that have been previously suggested and studied for avatars include realism \cite{inkpen_me_2011}, comfortableness using \cite{inkpen_me_2011},
comfortableness interacting with \cite{inkpen_me_2011}, appropriateness for work \cite{inkpen_me_2011}, creepiness \cite{inkpen_me_2011}, formality \cite{inkpen_me_2011}, resemblance to the person \cite{inkpen_me_2011}, trust \cite{ITU-P1320}, and emotion accuracy \cite{ITU-P1320}. We also include affinity to study the uncanny valley effect \cite{mori_uncanny_2012}.

The research questions we want to answer in this work are:

\begin{itemize}
    \item \textbf{RQ1}: Are the objective metrics currently used to develop avatars (PSNR, SSIM, LPIPS, FID, FVD) sufficient to achieve the performance goals of avatars, especially the human usability factors for avatars?
    \item \textbf{RQ2}: Which human usability factors are the most important for photorealistic avatars?
    \item \textbf{RQ3}: Can we develop an accurate and reproducible test framework to measure human usability factors for avatars?
    \item \textbf{RQ4}: Is there an uncanny valley effect for photorealistic avatars in the telecommunication scenario?
    \item \textbf{RQ5}: What avatar performance is needed to have the same level of performance as real video?
\end{itemize}

Our contributions in this work are:

\begin{itemize}
    \item We provide an open source test framework to subjectively measure photorealistic avatar quality of experience in ten dimensions using crowdsourcing.
    \item The crowdsourced subjective test framework is highly reproducible and accurate compared to a panel of experts. 
    \item We show that the correlation of nine of these subjective dimensions to PSNR, SSIM, LPIPS, FID, and FVD is weak and one (emotion accuracy) is moderate, motivating the need for an available subjective test framework as well as improved objective metrics to measure avatar performance.
    \item We analyze a wide range of avatars from photorealistic to cartoon-like and show some photorealistic avatars are approaching real video quality based on these subjective metrics. 
    \item We show that for avatars with a realism $>$ 2 (out of a 1-5 Likert scale) eight of the ten measured dimensions are strongly correlated, which leads to a dimensionality reduction of ten to three for the survey. 
    \item We show that for photorealistic avatars there is a linear relationship between avatar affinity and realism. In other words, there is no uncanny valley effect for photorealistic avatars in the telecommunication scenario; the more realistic the avatar is, the more affinity there is to the avatar. 
    \item We show that avatars that are not as realistic as real video will have lower trust, comfortableness using, comfortableness interacting with, appropriateness for work, formality, and affinity, and higher creepiness compared to real video.
\end{itemize}

In Section \ref{sec:related_work}, we review related work in this area. In Section \ref{sec:test_framework}, we describe the test framework design to measure avatar quality of experience, and in Section \ref{sec:survey_validation}, we show that the test framework is both reproducible and accurate. Using the test framework, we show the results and analysis in Section \ref{sec:results_analysis}. Finally, we provide conclusions, future extensions, system design implications, and limitations in Section \ref{sec:conclusions}.

\section{Related work}
\label{sec:related_work}

\subsection{Uncanny valley effect}
The famous uncanny valley effect was first hypothesized by Mori in 1970 and later translated by MacDorman and Kageki \cite{mori_uncanny_2012}, who described the phenomenon where humanoid objects that appear almost, but not quite, human evoke feelings of eeriness and discomfort in observers. Mori suggested when the affinity of humanoid objects is plotted on the y-axis against their human likeness on the x-axis, the affinity will increase as the likeness increases but then dips before going back up. This local dip in affinity is called the uncanny valley. For moving objects, Mori gave examples in order of human likeness: an industry robot, a humanoid robot, a zombie, a myoelectric hand, a bunraku puppet, an ill person, and a healthy person. The bottom of the uncanny valley occurred with the zombie and myoelectric hand examples. A meta-analysis of 72 experiments done on the uncanny valley effect is described in Diel et al.~\cite{diel_meta-analysis_2022}, which used a three-level meta-analysis model and revealed the uncanny valley effect had a large effect size. Ho et al.~\cite{ho_measuring_2017} and Mathur et al.~\cite{mathur_navigating_2016} were able to empirically reproduce the uncanny valley. Their studies included humanoid robots and the dip in affinity corresponded to images of robots. Tinwell \cite{tinwell_perception_2013} explores how avatar realism and facial expressions impact perceptions of eeriness and creepiness. Their results revealed that virtual characters that showed a lack of a startled response to a screaming sound were regarded as the most uncanny. Katsyri et al~\cite{katsyri_virtual_2019} studied the uncanny valley using carefully matched images of virtual faces varying from artificial to realistic using both painted and computer-generated images. The painted images showed a linear relationship between affinity and human likeness, while the computer-generated showed more of an uncanny slope rather (flattening of the curve) than an uncanny valley. 

\subsection{Measuring avatar quality of experience}
The ITU-T Rec.~P.1320 \cite{ITU-P1320} describes what is important for the quality of experience assessment of telemeetings with extended reality elements. This includes realism, trust, eye contact, body movement, facial expression, gestures, and proxemic cues. However, no measurement methodology is provided in this ITU-T recommendation. 

\subsubsection{Subjective evaluations of avatars}
Inkpen et al.~\cite{inkpen_me_2011} created a subjective test framework for avatars which included measurements of formality, non-creepiness, realism, resemblance, appropriateness for work, comfortable using, and comfortable interacting with. Our test framework significantly extends this work by adding measures for trust, affinity, and emotion accuracy (see Section \ref{sec:avatar_qoe}). We added state of the art photorealistic avatars and added real video clips as a baseline. We tested for multiple viewpoints and validated the test framework for both accuracy and reproducibility. We make the test framework open source to make it available to any researcher, and we use crowdsourcing which makes it fast and cost-effective.

Canales et al.~\cite{canales_impact_2024} conducted a study on the impact of avatar stylization and trust. Using the medical scenario with the avatar as the doctor, they measured trust, integrity, ability, and benevolence. While the style did not statistically significantly change these measurements, raters selected the mid-level avatar (between realistic and caricature) as their preferred doctor. However, this study only included the MetaHuman avatars and never measured the realism of these avatars. We found MetaHuman avatars have a realism mean opinion score (MOS)=2.3 (see Figure \ref{fig:MOS_Scores_Model_NoAngles}), which means the realistic avatars used in the study were not close to real video or SOTA photorealistic avatars. This may explain the different results reported by \cite{canales_impact_2024} and our present study, in which we show a strong relationship between realism and trust.

McDonnell et al.~\cite{mcdonnell_render_2012} rated ten levels of avatar stylization created by changing rendering settings, from ``toon pencil'' to highly realistic. The avatars were rated on appeal, familiarity, friendliness, and trustworthiness, which had a linear relationship to realism when the avatar was human-like (not cartoonish). The study did not include how realistic the avatars were compared to real video.

Bartneck et al. (2009) emphasize that avatars that better mimic human emotional cues tend to result in higher ratings of both emotional resonance and overall realism. Emotionally expressive avatars can create stronger feelings of empathy and presence in virtual environments (Garau et al., 2003).

Zell et al.~\cite{zell_stylize_2015} rated still images of avatars with different styles from cartoon to realistic and measured realism, appeal, reassurance, familiarity, attractiveness, and eeriness. The results show that realism is a poor predictor of appeal, eeriness, and attractiveness. This study did not include real images as a baseline, and the realistic images are not photorealistic.

Phadnis et al.~\cite{phadnis_avatars_2023} rated 5 different avatars on realism and acceptance in the context of business meetings. They showed that photorealism is a key attribute in selecting work avatars, though regional preferences and relationships with work colleagues using the avatar may also play a role. No real video was used as a baseline, and the realistic avatars do not look very photorealistic. 

Weidner et al.~\cite{weidner_systematic_2023} conducted a meta-review on 72 papers that compare various avatar representations. They concluded that the communication domain significantly benefits from realistic rendering styles of avatars. In particular, they believe that realistic visualizations are crucial in communication as they allow for non-verbal behavior, which is crucial to understanding the information that is being transmitted.

Gasch et al.~\cite{gasch_avatar_2024} examined how using avatars of varying realism in virtual reality video conferencing affects users’ sense of presence, privacy concerns, and acceptance of the technology for work and education. They find that hyper-realistic avatars generate a stronger sense of presence and higher user preference, although privacy concerns increase notably if another person uses someone else’s avatar.

Frampton-Clerk et al.~\cite{frampton-clerk_investigating_2022} explored how people perceive the realism of someone else’s look-alike avatar, focusing on key features like lip syncing, facial expression, and full body movement. They show that while full facial and body animations enhance realism, having only lip syncing can trigger unsettling (uncanny valley) reactions.

Garau et al.~\cite{garau_impact_2003} investigated how avatar realism and eye gaze control affect perceived communication quality in a shared immersive virtual environment. They show that matching higher avatar realism with a more realistic gaze model significantly enhances users’ sense of presence and face-to-face interaction, whereas lower-realism avatars do not benefit from advanced gaze behaviors.

Suk et al.~\cite{suk_influence_2023} studied how avatar facial appearance in immersive virtual reality (VR) affects users' perceived embodiment and presence. They find that avatars with facial features resembling the user's own face significantly enhance the sense of embodiment but have no substantial effect on the user's sense of presence.

Junuzovic et al.~\cite{junuzovic_see_2012} compared avatar, video, and audio conferencing for four-person workplace meetings and found that video conferencing was rated highest across all dimensions, including social presence, realism, and usefulness, due to superior non-verbal cues. Although avatars provided more sociability and a shared virtual space compared to audio alone, participants considered cartoon avatars inappropriate for professional settings and expressed frustrations with tracking inaccuracies, but noted avatar conferencing had potential if perfected.

Lin et al.~\cite{lin_visual_2023} explores the effectiveness of various visual indicators (such as color coding, scanning effects, transparency, and badges) to convey the authenticity of photorealistic avatars in social VR, and their impacts on perceived trustworthiness. They conclude that explicitly displaying the avatar user's name enhances perceived trust, while most other visual indicators negatively affect trust, particularly as interactivity and immersion levels increase.

Wang et al.~\cite{wang_real-and-present_2024} explores using life-size 2D video-based avatars in augmented reality (AR) teleconferencing on head-mounted displays (HMDs) to balance visual realism and the sense of co-presence. Through pilot studies and user evaluations, the authors identify optimal avatar placements for small-group conversations and demonstrate that 2D video avatars provide a superior balance between visual fidelity and co-presence compared to traditional methods, also highlighting how the Field of View (FoV) affects optimal avatar placement.

Yu et al.~\cite{yu_avatars_2021} compares two avatar embodiment techniques—point cloud reconstruction (PCR) avatars and virtual character-based avatars (3DVC)—in an asymmetric VR/AR teleconsultation system. They showed that despite occlusions and missing facial details, PCR avatars provided superior perceptions of copresence, social presence, behavioral realism, and humanness compared to virtual character avatars, although neither method significantly affected objective task performance.

\subsubsection{Subjective evaluation of video quality}
A recent review of subjective image and video quality assessment tools is given by Testolina et al.~\cite{testolina_review_2021}. ITU-T Rec.~P.910 \cite{itu-t_recommendation_p910_subjective_2021} provides a general subjective video quality assessment standard for multimedia applications. P.910 includes ACR (Absolute Category Rating), ACR-HR (ACR with Hidden Reference), DCR (Degradation Category Rating), and paired comparison (PC) methods, as well as rater qualifications, environment conditions, and video playback procedures. ITU-T Rec.~P.911 is a counterpart of P.910 but for audiovisual signals. ITU-T Rec.~P.912 \cite{itu-t_recommendation_p912_subjective_2016} provides a target-specific subjective video quality assessment standard, such as for faces, license plates, etc. ITU-T Rec.~P.913 \cite{ITU-P913} considers different displays and testing environments and provides flexibility on the rating scale and modality with mandatory reporting of test requirements \cite{pinson2014new}. Recently, ITU-T Rec. P.911 and P.913 are merged into the P.910 providing a single recommendation given their large overlapping parts. ITU-T Rec.~P.918 \cite{ITU-P918} details subjective assessment methods for five perceptual video quality dimensions, which can provide diagnostic information on the source of observed degradation. Finally, ITU-R BT.500 \cite{ITU-BT500} focuses on the video quality of broadcast television signals in a highly controlled environment.

Tominaga et al.~\cite{tominaga_performance_2010} conducted a comparison of eight different subjective video quality assessment methods and found that ACR was the most suitable for statistical reliability, assessment time, and ease of evaluation. 

Keimel et al.~\cite{keimel_qualitycrowd-framework_2012} describe an open source tool QualityCrowd that supports ACR video quality assessment. QualityCrowd is extended by Upenik et al.~\cite{upenik_large-scale_2021} to include a Double Stimulus Continuous Quality Scale (DSCQS). Rainer et al.~\cite{rainer_web_2013} describe the tool WESP, an open source tool that supports ACR, ACR-HR, DCR, and PC. 

Jung et al.~\cite{jung_isoiec_2021} provide a methodology to conduct remote subjective video quality assessment studies in which the raters download videos and view them manually on their devices. The methodology includes no tests for visual acuity, color blindness, environmental conditions, or hardware setup.

\subsubsection{Objective evaluations of avatars}
The most commonly used objective metrics for evaluating avatars are PSNR, SSIM, and LPIPS. Peak Signal-to-Noise Ratio (PSNR) calculates the ratio between the maximum possible power of a signal and the power of corrupting noise, measured in decibels (dB). It's commonly used to evaluate image and video quality but may not align well with human perception \cite{wang_image_2004}. Structural Similarity Index Measure (SSIM) assesses image similarity based on luminance, contrast, and structural information \cite{wang_multiscale_2003}. Learned Perceptual Image Patch Similarity (LPIPS) employs deep neural networks to model human perceptual similarity between images \cite{zhang_unreasonable_2018}.

Fréchet Inception Distance (FID) measures the similarity between the distribution of generated images and real images in the feature space of a pre-trained Inception network \cite{heusel_gans_2017}. Fréchet Video Distance (FVD) extends the concept of FID to video sequences, capturing both spatial and temporal information to evaluate the quality of generated videos \cite{unterthiner_fvd_2019}. Facial Landmark Distance (LMD) evaluates the accuracy of facial landmark predictions by measuring the Euclidean distance between the predicted and ground truth landmark positions, often normalized to account for scale variations \cite{sagonas_semi-automatic_2013}.

Chen et al.~\cite{chen_subjective_2024} provide a dataset of photorealistic avatar videos from 36 subjects with the following transmission degradations: temporal artifacts, reduced texture resolution, and reduced frame rate. Each video has 26 ratings using ITU-T Rec.~P.910 \cite{itu-t_recommendation_p910_subjective_2021} to measure the avatar quality in the presence of these impairments. A new objective metric (HoloQA), is provided to predict subjective quality based on these impairments. 

\subsection{Trust, empathy, and fatigue for telecommunication}
Nguyen et al.~\cite{nguyen_multiview_2007} showed when eye gaze is preserved in a teleconferencing system, the same level of trust can be achieved in remote meetings compared to face to face meetings. Nguyen et al.~\cite{nguyen_more_2009} also showed that when teleconferencing systems included hand gestures then the same level of empathy can be achieved in remote meetings compared to face to face meetings. 

Gale et al.~\cite{gale_eeg_1975} conducted a study monitoring EEG activity (a proxy to cognitive load) of 18 male subjects and found that arousal was highest when subjects gazed directly into the eyes of a male experimenter positioned 2 feet away. While EEG arousal diminished with increasing distance, it consistently remained higher during direct gaze compared to averted gaze at all distances tested. Bailenson et al.~\cite{bailenson_nonverbal_2021} discusses four primary reasons why video conferencing can be exhausting: excessive close-up eye contact, constantly seeing oneself on screen, reduced mobility, and increased cognitive load from interpreting nonverbal cues. 

\subsection{Virtual reality and kinetic embodiment}
Biehl et al.~\cite{biehl_not_2015} showed that local participants exhibited significantly more overlapping talk with remote participants who used an embodied proxy, than with remote participants in basic-video conferencing. In addition, while kinetic embodied technology increased the local participants’ perceived presence of remote teammates, it did not enhance remote participants’ own sense of telepresence.

Ratan et al.\cite{ratan_playing_2014} explores how people present themselves and perceive others in social VR, highlighting a strong preference among users to keep avatars consistent with their physical selves and a heightened importance of voice in verifying identity cues (such as gender, age, and ethnicity). It further shows that social VR can influence how users understand their own identities—especially for those exploring or affirming marginalized or changing self-images—through the immersive and embodied nature of virtual interactions.

Freeman et al.~\cite{freeman_body_2021} showed people interact socially in VR using many of the same verbal and nonverbal cues familiar from face-to-face (F2F) settings, but they deploy them differently: VR users rely more on direct verbal strategies (like immediately introducing a topic) and simpler gestures (like waving) than they do in F2F encounters. In contrast, F2F interactions more frequently involve subtle cues like moving closer or making small sounds to get someone’s attention, indicating that VR changes both the norms and practices of social engagement.

Toothman et al.~\cite{toothman_impact_2019} investigated how different types of avatar tracking errors in Virtual Reality—such as latency, vibration, popping, and stuttering—affect user experiences including performance, embodiment, enjoyment, usability, and spatial presence. While substantial tracking errors negatively impacted task performance and users' sense of embodiment, they did not significantly reduce users' social presence during virtual interactions, even at extreme error levels.

\begin{table*}[]
\begin{center}
\caption{Open source crowdsourcing video quality assessment systems}
\label{tab:comparison}
\scalebox{0.625}{
\begin{tabular}{|c|c|c|c|c|c|c|c|}
\hline
\textbf{Tool} & \textbf{Measures} & \textbf{Rater qualification} & \textbf{Viewing conditions} & \textbf{Hardware} & \textbf{Network} & \textbf{Accuracy} & \textbf{Reproducible} \\
\hline
QualityCrowd \cite{keimel_qualitycrowd-framework_2012} & ACR, DSCQS & N & N & N & N & Y & N \\
\hline
WESP \cite{rainer_web_2013} & ACR, ACR-HR, DCR, PC & N & N & N & N & N & N \\
\hline
avrateNG \cite{rao_towards_2021} & ACR & N & N & N & N & Y & N \\
\hline
Ours & ACR, ACR-HR, DCR, CCR & Y & Y & Y & Y & Y & Y \\
\hline
\end{tabular}
}
\end{center}
\end{table*}


\section{Test framework design}
\label{sec:test_framework}
Our test framework is based on our crowdsourcing implementation to measure video quality \cite{naderi_crowdsourcing_2024} 
(a more detailed description is given in \cite{naderi_crowdsourcing_2023}). 
We first describe that implementation, and then, in Section \ref{sec:avatar_qoe} how we extend it to the avatar quality of experience task. 

\subsection{Measuring video quality}
\label{sec:video_quality}
Measuring the quality of video is an important task in many engineering areas, such as video codec development, video enhancement, and video telecommunication systems. As a result, there are many standards and metrics for measuring video quality, though the gold standard is the subjective test done in a controlled lab environment. The most prevalent of these standardized subjective tests is the ITU-T P.910 \cite{itu-t_recommendation_p910_subjective_2021}. However, using P.910 in practice is slow due to the recruitment of test subjects and the limited number of test subjects, and expensive due to paying qualified test subjects and the cost of the test lab. The speed and cost result in the vast majority of research papers not using P.910 but rather objective functions that are not well correlated to subjective opinion. 

An alternative to lab-based subjective tests is to crowdsource the testing, and there are several such systems that do this (see Table \ref{tab:comparison}). However, none of these systems have the rater, environment conditions, and hardware qualifications that P.910 requires. In addition, these existing systems have not been rigorously validated to show they are accurate compared to a P.910 lab-based study and give reproducible results. We provide a crowdsourced implementation that includes rater, environment, hardware, and network qualifications, as well as gold and trapping questions to ensure quality. We include a validation study that shows it is both accurate and highly reproducible compared to existing P.910 lab studies. The tool is open sourced and can be used on the Amazon Mechanical Turk platform for wide-scale usage. The tool has been used in the CVPR 2022 CLIC and DCC 2024 CLIC challenges ({\url{http://compression.cc/}) 
to provide the challenge metric for machine learning-based video codecs.

\subsubsection{Implementation}
\label{sec:implementation}
The implementation at \url{https://github.com/microsoft/P.910} 
can be used either as an integrated survey in Amazon Mechanical Turk (MTurk) or as an external survey deployed on a dedicated Web server. We also provide a lightweight container-based web application that can serve the experiment. The web application can easily be deployed on any Linux virtual machine. As a result, this implementation can be used for both crowdsourcing tests or remote testing with a dedicated panel of participants.

The open source implementation includes ACR, ACR-HR, DCR, and CCR. All methods can be used with either a five or nine-point discrete Likert scale. We also followed and extended best practices on video and speech quality assessment in crowdsourcing \cite{hossfeld_best_2014, ITU-PSTR-CROWDS} in our implementation. The avatar quality of experience uses ACR and CCR measurements, and extensions could use ACR-HR and DCR.  

\subsubsection{Tools}
We provide a set of program scripts to ease the interaction with the system and avoid operation errors. The scripts are used to create trapping sequences, process the submitted answers, and interact with MTurk. Figure~\ref{fig_dfd} shows the data flow diagram of the system.

Trapping sequences are customized to the dataset under the test and created by adding a text message asking participants to select a specific score. A test configuration and URLs for test sequences, trapping sequences, gold, and training clips are provided to the master script, which creates the HTML template of the test, a list of variables, and a configuration file for the result parser. Gold clips are video sequences whose quality is known to the experimenter. A test can be created in the HIT App Server by providing the HTML template and a list of variables. A generic project description and list of URLs can be downloaded and used to create a new test in MTurk.

The submitted answers are provided to the result parser script along with the configuration file. The script performs data cleansing and aggregates the valid and reliable ratings over the test sequence and over the test condition (i.e., Hypothetical Reference Circuits - HRCs). Reports on the list of bonus assignments and lists of the submissions to be accepted/rejected or extended are also generated.

\begin{figure}[htbp]
\centerline{\includegraphics[width=\columnwidth]{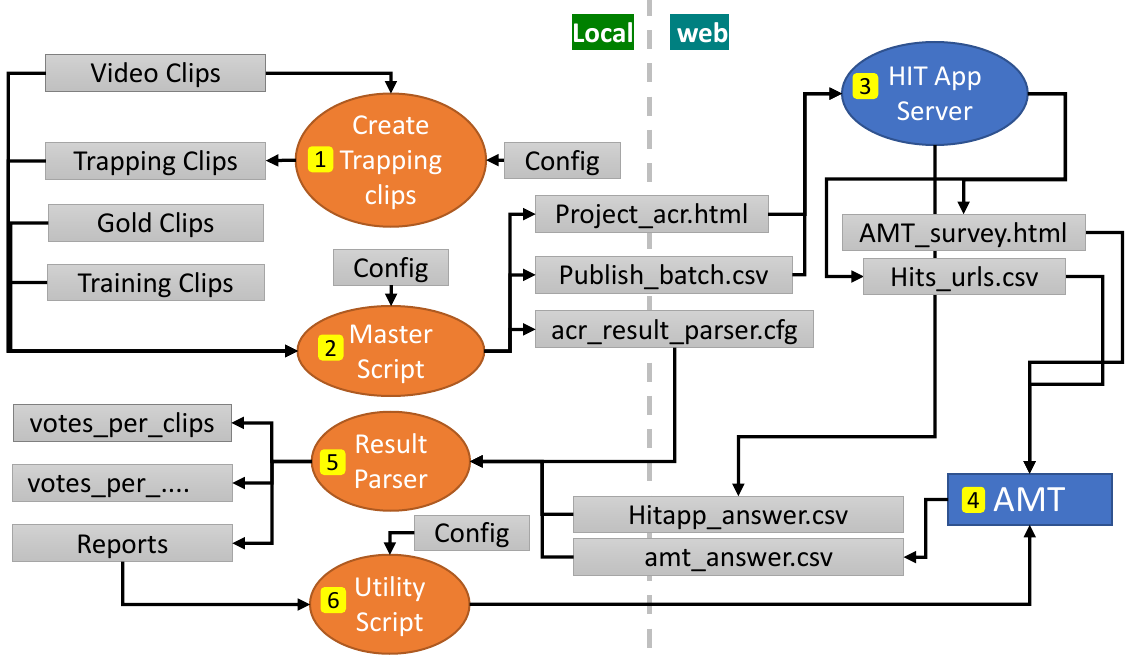}}
\caption{Data Flow Diagram. }
\label{fig_dfd}
\end{figure}

\subsubsection{Test components}
The subjective test is organized in different sections from the participant's perspective (see Figure~\ref{fig_info}). Each section is designed to instruct the test participant, qualify the participant, their environment, and their setup, and collect their votes. The \textit{instructions} and \textit{ratings} sections are included in all tests, whereas the \textit{qualification}, and \textit{calibration} only need to be performed once per test. The \textit{setup} and \textit{training} sections are shown  (e.g., once per hour).

\begin{figure}[htbp]
\centerline{\includegraphics[width=\columnwidth]{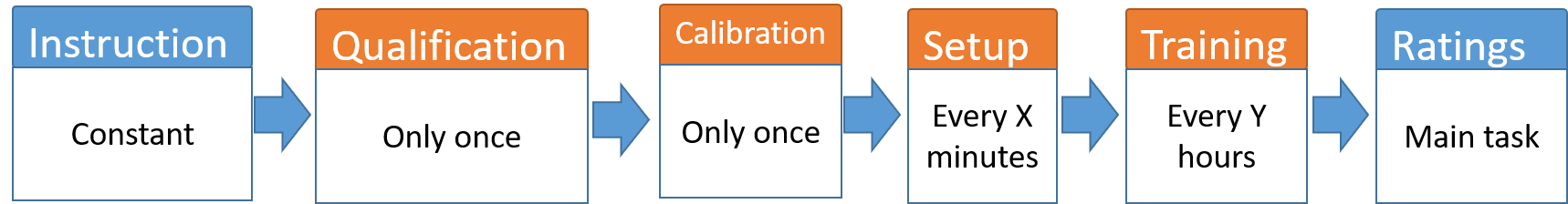}}
\caption{The crowdsourcing test from the participant's perspective.}
\label{fig_info}
\end{figure}

A set of automatic measurements are performed on the test's loading time. The experimenter can restrict test participants to specific viewing devices (i.e., mobile, PC, or both), minimum screen refresh rate, and minimum resolution.

\subsubsection{Video playback}
We developed an HTML5 video playback component that downloads all videos to the browser's local storage to avoid network latency. Videos are played in full-screen mode and participants must watch the entire video before voting. The player records the playback duration, which is used in the data cleansing process. The experimenter can choose to present videos in their original size or scaled to fill the participant's viewing device. For DCR or CCR tests, the reference and processed video sequences are shown sequentially with a one-second gray screen in between.

\subsubsection{Qualification}
Within the qualification section, the eligibility of test participants is evaluated. According to P.910, participants should be screened for normal color vision and normal or corrected-to-normal visual acuity (i.e., no error on the 20/30 line of a standard eye chart) \cite{ITU-P910}.
The standard Ishihara test for color vision \cite{clark1924ishihara} includes 15 plates in the normal and 6 plates in the short version, which are both too long for a crowdsourcing test. In a prestudy, we invited 300 participants from MTurk (34\% male) and 191 participants (91\% male) from two internet communities dealing with color vision deficiencies. Both groups participated in the full Ishihara test in which 6\% of participants from MTurk and 96\% from the online forums were detected as color blind. Applying a decision tree classifier with entropy as the criterion revealed that only using Plates 3 and 4 reached 98\% accuracy (sensitivity 0.996, specificity 0.95). Consequently, we only use these two plates in the qualification section.

We use the Landolt ring optotypes to measure visual acuity, as recommended by ISO 8596:2017 \cite{iso_8596}. The participant is presented with broken rings (like the letter C) with a gap in 8 directions. The diameter of the ring is 5 times the size of the gap. The visual acuity is the inverse value of the gap size (in arc minutes) of the smallest identified Landolt ring. In each row (i.e., ring size), 5 samples are presented and the participant must answer 3 or more correctly to pass that size. Our implementation of the visual acuity test consists of two steps: setup and answering up to 5 Landolt rings at a specific size. In the setup section, the participant is asked to adjust the size of a given picture (here a credit card) on their screen until it is the same size as a real credit card. This is used to estimate the size of a pixel on their screen. The participant is then asked to sit in the range of 50 to 75 cm from the screen. The corresponding Landolt ring size is calculated, and the participant must correctly identify the direction of 3 out of 5 rings at that size to pass the test.

\subsubsection{Display calibration and instructions}
Participants are asked to set the resolution of their device to default or recommended value suggested by their operating system. We also ask them to perform display color calibration using methods provided by their operating system. This section provides information on how to perform these tasks for Windows and Mac operating systems and will only be shown once during the test.

In the instruction section, a sample video for some of the perceptual quality dimensions~\cite{schiffner2017defining, ITU-P918} (i.e., fragmentation, discontinuity, and uncleanness) is presented to the participants. They are also informed that impairments can happen in a specific area or time within a video clip and the impairments are not limited to the presented samples.

\subsubsection{Setup}
\label{setup}
The setup section evaluates the viewing conditions, including the brightness of the screen, room light, and viewing distance. A brightness/light calibration task is used, in which participants count the number of geometric shapes in a picture. If they get the answer wrong, they are encouraged to adjust their screen brightness, and lighting, or repeat the task. The picture is a matrix of 4x4 squares, with each square having a different gray background and a triangle or circle in different sizes and locations. The foreground color of the shapes is close to the background color of the square.

The crowdsourcing test also includes a short paired-comparison task to evaluate participants' viewing distance. This task is inspired by \cite{naderi2020application}, which is the recommended method to evaluate crowd workers' setup and environment in a speech quality assessment test \cite{ITU-P808}. In this task, three pairs of images are presented to the participants, who are asked to select the image of better quality. One of the images in each pair is distorted with a blur effect. The blur effects are selected based on \cite{treutwein1995adaptive}, which found that participants can correctly distinguish between blurred images at different distances.

\subsubsection{Training and rating}
The training section includes videos from the training set, which cover the entire range of the rating scale. It also includes a trapping clip to test the participant's understanding. Participants are alerted if they provide a wrong answer and asked to watch the video again. The training section is shown periodically to keep its anchoring effect, following best practices in the speech quality domain \cite{naderi_open_2020}.

The rating section includes ten video clips, one trapping clip, and one gold clip. The trapping and gold clips are inserted automatically and used in the data cleansing step in post-processing.

\subsubsection{Validation}
\label{sec:validation}
We used the VQEG HDTV datasets \cite{video2010report} to evaluate the validity and reliability of our implementation. 
The datasets contain coding only and coding plus transmission error impairments. They were created to validate objective video quality models that predicted the quality of High Definition Television (HDTV). The video materials and subjective data from experiments VQEGHD1-5 (in a laboratory) are made publicly available in the Consumer Digital Video Library \url{https://www.cdvl.org}. Each experiment includes 168 video clips, with 24 shared between all experiments. Each experiment includes its specific dataset where 21 HRCs were applied to 9 source videos (i.e., 144 processed sequences). We used the datasets from experiments VQEGHD3 and VQEGHD5 for which 40" (native resolution 1920x1080) and 24" (native resolution 1920x1200) displays were used in the laboratory viewing sessions by 24 test participants, respectively. The test sequences have various bit rates; 1-15 Mbps for VQEGHD3 and 2-16 Mbps for VQEGHD5. The test sequences created from two source clips in the dataset VQEGHD5 were not included in the published package, leading to 136 sequences. Results of laboratory-based subjective tests were published in \cite{video2010report} by VQEG.

We conducted six crowdsourcing studies with the two above-mentioned datasets. In one of the tests, the VQEGHD5 dataset was used. In the rest of the five tests, we used the VQEGHD3 dataset on five separate days, each with unique raters, to evaluate the reproducibility of our implementation.
In all tests, participants with a minimum display resolution of 1280x720 and a refresh rate of 30Hz were allowed.

\subsubsection{Accuracy}
In each experiment, 10 test sequences, one gold clip, and one trapping clip were presented in one session and we aimed to collect 30 ratings per clip. In the experiments using VQEGHD3 an average of 71\% submissions passed the data cleansing step; the rest were not used due to the rater providing a wrong answer to the gold question, longer playback duration, failure on the second brightness check, low variance in ratings, or a wrong verification code. On average we had 21 accepted votes per test sequence with a minimum of 15 ratings. In the experiment using VQEGHD5, 78.8\% of submissions passed the data cleansing step. On average we had 25 accepted votes per test sequence with a minimum of 20 ratings. The results are presented in Table~\ref{tab:validity} per test sequences and in Table~\ref{tab:validity_cond} per HRCs. We observed a strong correlation between laboratory and crowdsourcing subjective tests ($\overline{r} = 0.952$ per test sequence and $0.964$ per HRC).

\begin{table}[tb]
    \caption{Comparison between laboratory and crowdsourcing experiments (sequence level).}
    \label{tab:validity} 
    \begin{center}
    \resizebox{0.8\columnwidth}{!}{%
        \begin{tabular}{ l c c  c  c  c  c }
        \toprule
        \textbf{Dataset} &	 \multicolumn{3}{c}{\textbf{MOS}} &
         \multicolumn{3}{c}{\textbf{DMOS}} \\
        
        & {\small \textbf{PCC}}&	{\small\textbf{SPCC}}&	{\small\textbf{RMSE FOM}} & {\small \textbf{PCC}}&	{\small\textbf{SPCC}}&	{\small\textbf{RMSE FOM}} \\ 
        \midrule
        VQEG HDTV3  -run1 & 0.956 & 0.949 & 0.333 & 0.948 & 0.949 & 0.362\\
        VQEG HDTV3  -run2 & 0.964 & 0.951 & 0.302 & 0.946 & 0.939 & 0.370\\
        VQEG HDTV3  -run3 & 0.959 & 0.949 & 0.323 & 0.940 & 0.942 & 0.389\\
        VQEG HDTV3  -run4 & 0.917 & 0.913 & 0.455 & 0.904 & 0.922 & 0.489\\
        VQEG HDTV3  -run5 & 0.947 & 0.923 & 0.367 & 0.932 & 0.909 & 0.415\\
        VQEG HDTV5        &	0.970&	0.957 &	0.278 & 0.965 & 0.958 & 0.299\\ 
        \bottomrule
        \end{tabular}
    }
    \end{center}
\end{table}

\begin{table}[htbp]
    \caption{Comparison between laboratory and crowdsourcing tests in HRC level
    }
    
    \label{tab:validity_cond} 
    \begin{center}
    \resizebox{0.6\columnwidth}{!}{%
        \begin{tabular}{ l c c  c  c }
        \toprule
        \textbf{Dataset} &	 \multicolumn{3}{c}{\textbf{MOS}} \\

        & {\small \textbf{PCC}}&	{\small\textbf{SPCC}}&	{\small\textbf{RMSE}} & {\small\textbf{RMSE FOM}}  \\ 
        \midrule
            VQEG HDTV3 -run1 &0.967	& 0.980	& 0.655	& 0.248\\
            VQEG HDTV3 -run2 &0.977	& 0.982	& 0.618	& 0.211\\
            VQEG HDTV3 -run3 &0.968	& 0.981	& 0.577	& 0.245\\
            VQEG HDTV3 -run4 &0.940	& 0.975	& 0.706	& 0.333\\
            VQEG HDTV3 -run5 &0.965	& 0.972	& 0.671	& 0.257\\
        \bottomrule
        \end{tabular}
    }
    \end{center}
\end{table}

\subsubsection{Reproducibility}
 On average 63 unique workers participated in each run. We observed a $PCC=0.971$ and $SRCC=0.95$ on average between the MOS values of sequences in the five different runs. The correlation coefficients between the runs are reported in Table~\ref{tab:repro}.
We also fitted a linear mixed-effects model (LMEMs) with random intercept in which test sequences and runs are considered as fixed factors and participants as a random factor. The result shows there was no significant effect of runs on the subjective ratings ($\chi^2(4)=2.691, p=0.611$).

\begin{table}[tb]
    \caption{Correlation coefficients between five runs of the VQEGHD3 dataset. Pearson correlation coefficients are on the upper triangle and Spearman's rank correlation coefficients are on the lower triangle.}
    \label{tab:repro} 
    
    \begin{center}
    \resizebox{0.5\columnwidth}{!}{%
        \begin{tabular}{ c c c  c  c  c }
        \toprule
        &\textbf{Run 1} &\textbf{Run 2} &\textbf{Run 3} &\textbf{Run 4} &\textbf{Run 5} \\ 
        \midrule
        \textbf{Run 1} &         &0.984  &0.987  &0.957   &    0.977\\
        \textbf{Run 2} &0.959    &       &0.985  &0.957   &    0.977\\
        \textbf{Run 3} &0.974    & 0.969 &       &0.952   &    0.972\\
        \textbf{Run 4} &0.943    & 0.941 & 0.942 &        &    0.956\\
        \textbf{Run 5} &0.954    &0.947  & 0.942& 0.933   &    \\
        
        \bottomrule
        \end{tabular}
    }
    \end{center}
\end{table}

\subsection{Avatar quality of experience survey design}
\label{sec:avatar_qoe}
The goal of the photorealistic avatar survey is to measure the  quality of experience when using the avatar in multiple dimensions. We include realism, trust, and facial expressions (emotion accuracy) as suggested in ITU-T Rec.~P.1320 \cite{ITU-P1320} to measure the quality of experience. We also include formality, creepiness, resemblance, appropriateness for work, comfortable using,
and comfortable interacting with as used in \cite{inkpen_me_2011}. We added affinity to study the uncanny valley effect, which is the simplest dimension that has been used to validate the uncanny valley \cite{mathur_navigating_2016} and is what is used in the original hypothesis of the uncanny valley \cite{mori_uncanny_2012}.  

We used each participant's eligibility tests, environment, setup, and device checks used in subjective video quality assessment in crowdsourcing described in Section \ref{sec:video_quality} in our survey. Participants are screened for normal color vision, normal or corrected-to-normal visual acuity, and their network speeds are tested.  

Upon passing the qualification tests, a training session with descriptive feedback is provided to ensure that participants are exposed to various types of video representations in online meetings. We provide two survey templates: Template A focuses on the avatar representation itself, while Template B compares the avatar and the original video, focusing on resemblance to the person and accurate emotion representation. Figure~\ref{fig:templates} illustrates the survey items and their representations for both templates. Multiple statements are provided and participants should specify how much they agree with each item in a five-point discrete scale. Participants can begin rating after watching the video in full-screen mode. The view is looped back to a smaller screen size during the rating process. For Template B, an original video of a person is horizontally stacked with the avatar video to make sure participants can assess how well a person and their emotions is represented by the avatar.

To ensure the reliability of the ratings, we have included three mechanisms. \textit{A) Golden clips:} For these, we already know the correct answer for one or more items in the survey. One golden clip is added to a set of ten real clips, which are assessed by a participant in a single session. A significant deviation from the expected answers will result in the discarding of all answers in that session. Examples of golden clips could be a real video or a completely cartoonish avatar for Template A and a video of a person combined with an avatar of another  person for Template B. \textit{B) Trapping items:} For each video, a trapping item is randomly selected and inserted into the item list. The answer to the trapping item is obvious and irrelevant to the video if the participant reads the statement carefully. These items are used to catch random clickers or straight liners. An example is: \textit{I cannot read text in English}. \textit{C) Repeated items:} One of the items in the list is repeated. We expect to receive the same answer each time. This helps us check the consistency of the provided answers. In Figure~\ref{fig:template_A}, the item in position three is repeated in position eight. The trapping and repeated items are only used in Template A as they could not be useful in Template B with only two original items. For Template B, an additional mechanism called \textit{trapping clips} is used. For these clips, the video is interrupted in the middle by a message on the screen asking participants to select a specific answer to demonstrate their attention. Since the beginning and end of trapping clips are completely similar to other clips, participants need to pay attention throughout the entire video to catch these instructions. Similar to other mechanisms, failures lead to the rejection of the entire submission.

Finally, in post processing, besides the answers to the above mentioned tests, outliers and variance in ratings will be checked. The accepted submissions are aggregated per clip and per model, and MOS, standard deviation, and 95\% confidence interval for each item will be reported.

\begin{figure*}
    \centering
    \subfloat[]{\includegraphics[width=\columnwidth]{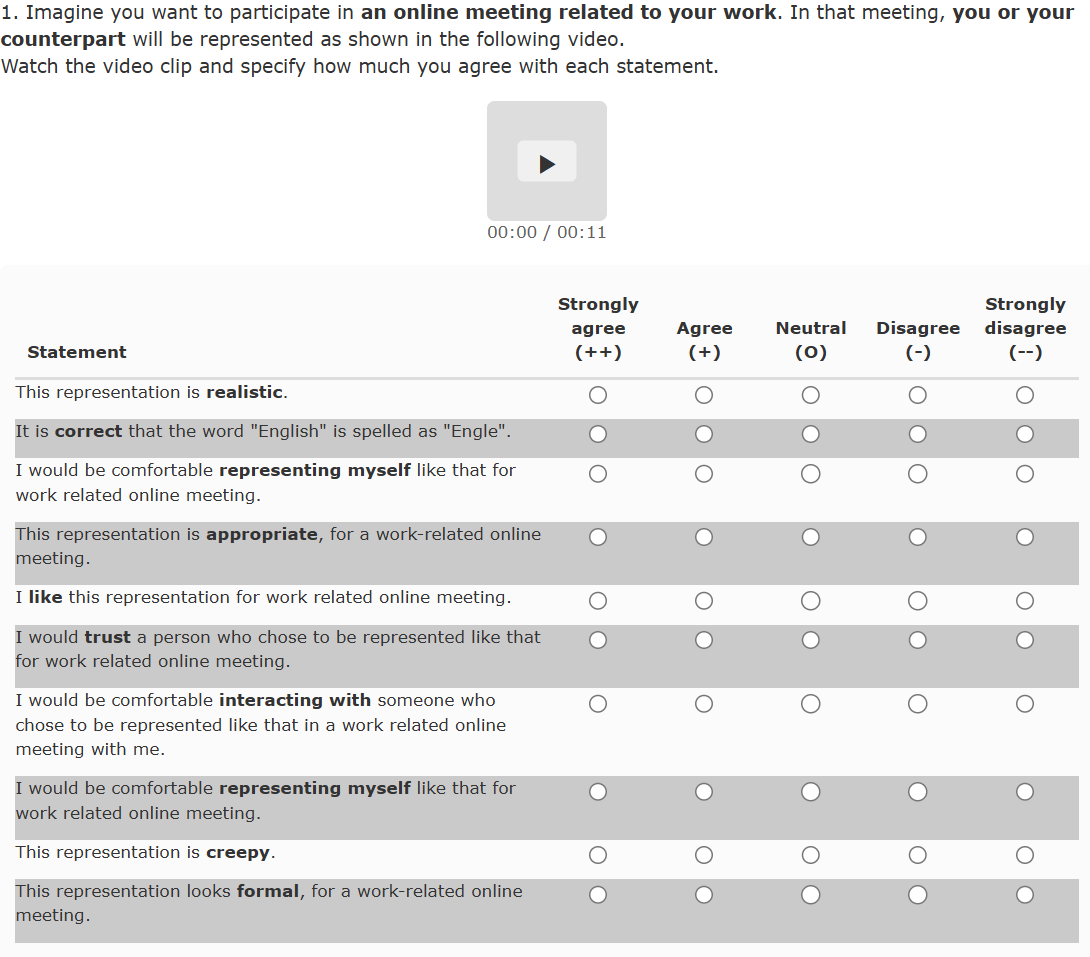}
    \label{fig:template_A}
    }
    
    \subfloat[]{\includegraphics[width=\columnwidth]{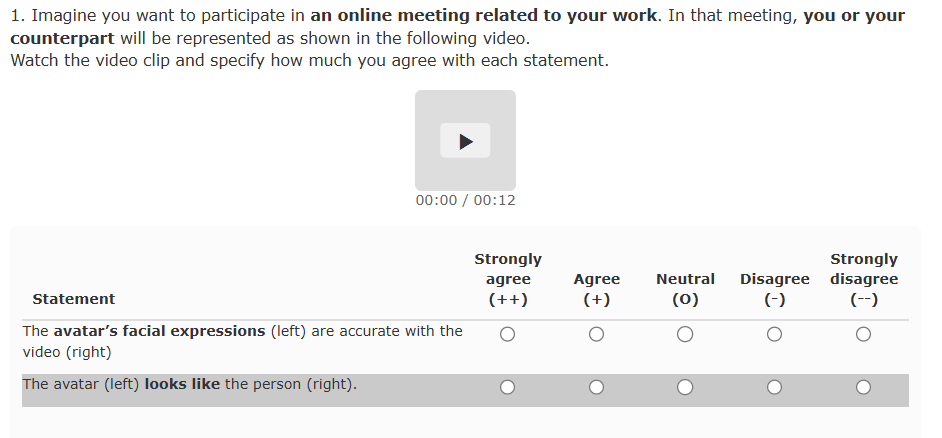}%
    \label{fig:template_B}
    }
        
    \caption{Items in the survey. (a) Template A as represented in the survey including a trapping and repeated items, (b) Template B focuses on resemblance to the person and emotion accuracy.}
    \label{fig:templates}
\end{figure*}

\section{Survey validation}
\label{sec:survey_validation}
\begin{figure*}
    \centering
    \includegraphics[width=1\linewidth]{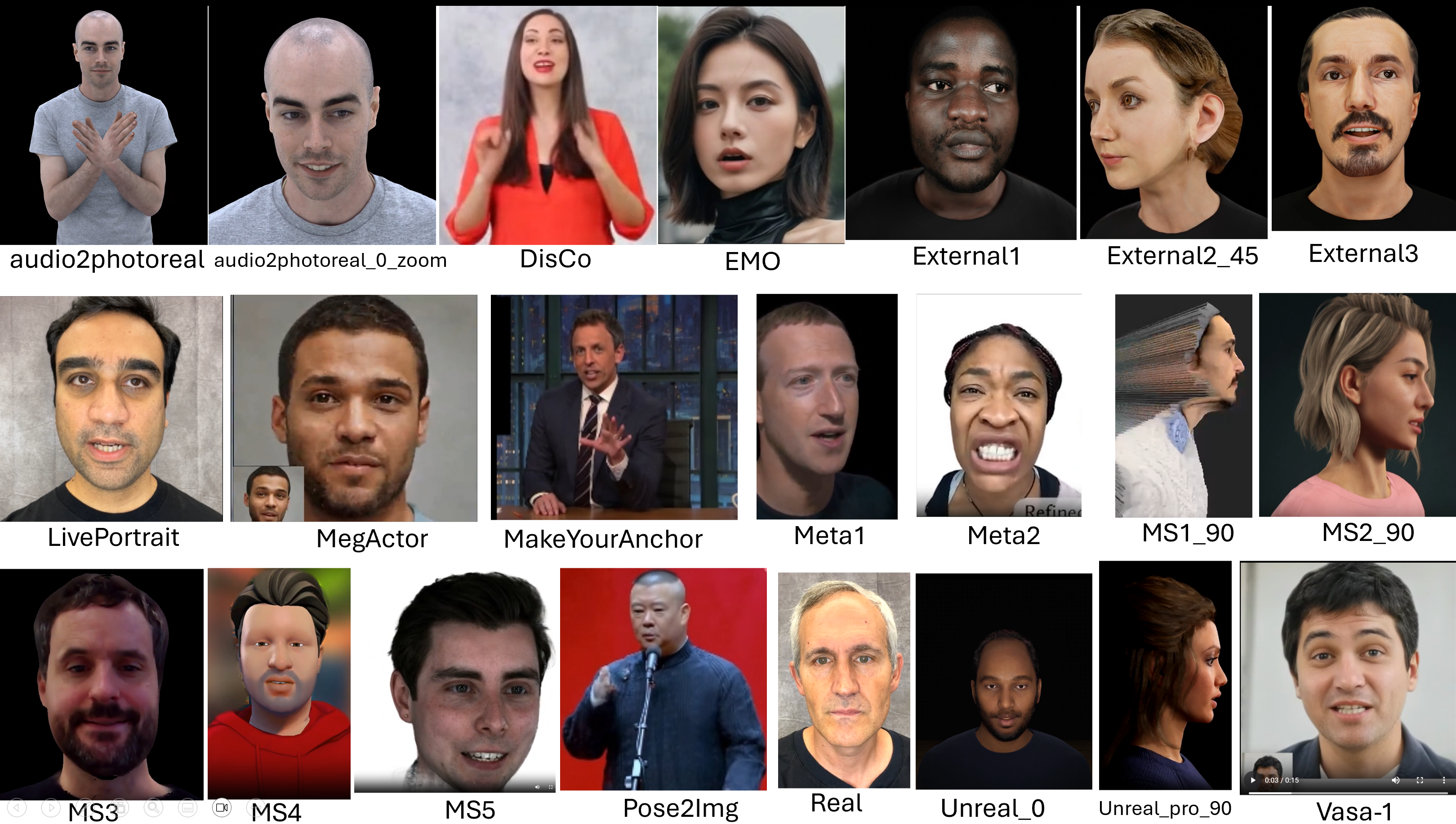}
    \caption{Avatars used for the survey}
    \label{fig:AvatarsUsed}
\end{figure*}
\subsection{Survey comparisons}
\label{sec:survey_comparisons}
In order to ensure accuracy and to validate the correlations across items, we tested out 3 different versions of the survey.
\begin{itemize}
    \item The first version is as shown in Figure \ref{fig:template_A}. This version reduces the time needed to complete the survey and the raters need only to look at the clip once before answering all of the questions.
    \item The second version broke each of the items into separate tabs, requiring the rater to switch tabs to answer a single item. The idea was to force the rater to rethink the answer on every item and minimize the effect of straight lining. This format takes longer than the first format due to extra tab navigation.
    \item The third version of the survey built on the idea of the second one and randomized the clip and the item on each tab. This format takes significantly longer to complete due to the need to play the clips 8 times more than the first format.
\end{itemize}
Across the surveys, we consistently saw a good correlation amongst the items. The correlation only dipped marginally for the third format as compared to the first format.

\subsection{Test set}
To test the test framework we created a test set consisting of 237 video clips of 19 different avatar models. For Template A the video clips were of people talking. For Template B the video clips were also include facial emotions: happy, sad, surprised, fear, anger, and disgust. We found that applying Template A to the facial emotion clips resulted in all the clips being rated as creepy, including the real video clips. 

We used 19 different avatars in the test set. The VASA-1 \cite{xu_vasa-1_2024}, MS1, MS2, MS3, MS4 and MS5 avatars are from Microsoft.
External1, External2, and External3 are 3 variations from the same organization. External1 is the AI model, External2 is the computer graphics based avatar and External3 is the set of avatars we created from the iPhone application from the organization. The Unreal\footnote{\url{https://www.unrealengine.com/en-US/metahuman}} avatars are models created using Epic's Metahuman avatar creator. The Unreal\_pro avatars are animations of Epic's Metahuman store models created by professionals, while the non-pro Unreal avatars are created by the authors. The EMO \cite{tian_emo_2024}, MegActor \cite{yang_megactor_2024}, Make-Your-Anchor \cite{huang_make-your-anchor_2024}, DisCo \cite{wang_disco_2024}, Pose2Img \cite{qian_speech_2021}, and Meta2 \cite{cao_authentic_2022} avatars were scraped from their respective project pages. The Zoom avatar was a screen capture of the avatar in Zoom's video conferencing application. The Meta1 avatar is from the Mark Zuckerberg, Lex Fridman interview\footnote{\url{https://www.youtube.com/watch?v=MVYrJJNdrEg}}. Audio2Photoreal and LivePortrait \cite{guo_liveportrait_2024} were created using their open sourced models on GitHub. 9 of the 19 avatar models have more than one viewpoint variation (frontal, $45^\circ$ off-center and $90^\circ$ off-center). Thus we have a total of 40 different avatar conditions with an average of 5.4 clips per variation for Template A. This also includes the baseline clips of real recordings of the actors. For Template B, we only used models that had input video clips available, resulting in 26 different avatar conditions from 13 different avatars with an average of 5 clips per variation. Table \ref{tab:test_set} provides the number of people for each avatar, the number viewpoints, whether speech is included, and if there was a reference video driving the avatar. 

This test set should help answer RQ1, as we can compute objective and subjective metrics on it and compute their correlation. We can also use this test set to answer RQ2, as we can determine the importance and redundancy of each dimension measured in the survey. RQ3 can be answered by using this test set to show the results are both accurate compared to experts and reproducible. By analyzing affinity and realism from this test set we can determine if we see a dip in affinity as realism increases, which answers RQ4. Finally, because we include real videos in this test set (which we consider as the baseline), we can determine the gaps between real video and avatars in each dimension.

\begin{table}[t]
\caption{Details of the avatar test set. Races are Caucasian, Asian, Black, and Genders are Male/Female.}
\label{tab:test_set} 
\begin{center}
\resizebox{0.9\columnwidth}{!}{%
    \begin{tabular}{l c c c c c c}
    \toprule        
    \textbf{Avatar} & \textbf{\# people} & \textbf{Race C/A/B} & \textbf{Gender M/F} & \textbf{\# viewpoints} & \textbf{Speech} & \textbf{Reference video} \\
    \midrule
    Audio2Photoreal & 4 & 3/0/1 & 2/2 & 5 & Y & N \\
    DisCo & 4 & 2/2/0 & 3/1 & 1 & N & N \\
    EMO & 2 & 1/1/0 & 1/1 & 1 & Y & Y \\
    External1 & 4 & 3/0/1 & 3/1 & 3 & Y & Y \\
    External2 & 4 & 3/0/1 & 3/1 & 3 & Y & Y \\
    External3 & 4 & 2/2/0 & 4/0 & 3 & Y & Y \\
    LivePortrait & 4 & 2/2/0 & 4/0 & 1 & Y & Y \\
    MegActor & 4 & 2/1/1 & 2/2 & 1 & N & N \\
    Make-Your-Anchor & 6 & 3/3/0 & 5/1 & 1 & N & N\\
    Meta1 & 2 & 2/0/0 & 2/0 & 1 & Y & N \\
    Meta2 & 6 & 3/2/1 & 4/2 & 1 & N & Y \\
    MS1 & 4 & 2/2/0 & 4/0 & 3 & Y & Y \\
    MS2 & 5 & 3/2/0 & 3/2 & 3 & Y & N \\
    MS3 & 5 & 4/1/0 & 4/1 & 4 & Y & Y \\
    MS4 & 4 & 1/3/0 & 4/0 & 1 & Y & N \\
    MS5 & 2 & 2/0/0 & 2/0 & 1 & Y & Y \\
    Pose2Img & 4 & 2/2/0 & 3/1 & 1 & N & N \\
    Unreal & 4 & 2/2/0 & 4/0 & 3 & Y & Y \\
    Unreal\_pro & 10 & 5/1/4 & 7/3 & 3 & Y & N \\
    VASA-1 & 6 & 4/1/1 & 3/3 & 1 & Y & N\\
    \midrule
    Total & 88 & 51/27/10 & 67/21 & & & \\
    \end{tabular}
}
\end{center}
\end{table}

\subsection{Reproducibility}
\label{sec:repro}
To test the reproducibility from run to run, for each template we repeated our crowdsourcing test five times with a mutually exclusive group of workers, on separate days on Amazon Mechanical Turk. On average we have collected 22.8 votes per video clip using Template A and 22.4 votes using Template B in each run. We calculated the MOS per clip and per condition (avatar variation) and show the correlation between different runs and scales in Tables \ref{tab:ReproTablePerCondition} and \ref{tab:ReproTablePerClip}. The results show a strong correlation between different runs at the clip (average PCC=0.968) and condition levels (average PCC=0.987) and confirm the survey provides reproducible scores.

\subsection{Accuracy}
To test the accuracy of the survey, we compared the results of a panel of experts with crowdsourced raters. We randomly selected 10 different avatar models from Section~\ref{sec:repro} for the laboratory based expert viewing of Template A and B. Four video clips from each avatar were included in the test and five experts rated them. The correlation coefficients between expert viewing scores and subjective ratings from crowdsourcing for this subset are reported in Table~\ref{tab:expert_vs_cs}. The results show a strong correlation coefficient between expert ratings and subjective scores collected using this survey (average PCC=0.904).

\begin{table}[t]
\caption{Pearson correlation coefficients (PCC) and Spearman rank correlation coefficients (SRCC) between subjective scores from expert viewing and crowdsourcing test in model level.}
\label{tab:expert_vs_cs} 
\begin{center}
\resizebox{0.4\columnwidth}{!}{%
    \begin{tabular}{l c c }
    \toprule        
    \textbf{Items} & 
    \textbf{PCC } & 
    \textbf{SRCC} \\
    
    \midrule
    \textbf{Appropriate} &  0.929 & 0.900  \\
    \textbf{Comfortable Interacting} & 0.944 & 0.927  \\
    \textbf{Comfortable Using} & 0.951 & 0.927 \\
    \textbf{Formal} & 0.902 & 0.891 \\
    \textbf{Affinity} & 0.935  & 0.924 \\
    \textbf{Not Creepy} & 0.940  & 0.936 \\
    \textbf{Realistic} & 0.880  & 0.927 \\
    \textbf{Trust} & 0.882  & 0.912  \\
     \midrule

    \textbf{Emotion accuracy}  & 0.862 & 0.976 \\
    \textbf{Resemblance} &  0.817 & 0.915 \\
    \midrule
    \textbf{Average}  &  0.904 & 0.928  \\
    \midrule
    \end{tabular}%
}
\end{center}
\end{table}


\section{Results and analysis}
\label{sec:results_analysis}
\begin{table*}[htb]
    \centering
    \caption{Pearson correlation between different runs of the reproducibility test in model level.} 
    \resizebox{\columnwidth}{!}{%
    \begin{tabular}{c | c c c c c c c c | c c}
         \toprule
         & \multicolumn{8}{c}{\textbf{Template A}} &\multicolumn{2}{|c}{\textbf{Template B}} \\
         & & \textbf{Comfortable} & \textbf{Comfortable} & & & & & &\textbf{Emotion} &\\
          & \textbf{Appropriate} & \textbf{Interacting} & \textbf{Using} & \textbf{Formal} & \textbf{Affinity} & \textbf{Not Creepy} & \textbf{Realistic} & \textbf{Trust} & \textbf{accuracy} &\textbf{Resemblance}\\          
         \midrule
         \textbf{Run1 - Run2} & 0.983 & 0.982 & 0.983 & 0.985 & 0.982 & 0.980 & 0.983 & 0.981 & 0.993 & 0.988\\
         
         \textbf{Run1 - Run3} & 0.991 & 0.992 & 0.992 & 0.993 & 0.991 & 0.987 & 0.993 & 0.990  & 0.994 & 0.978\\
         
         \textbf{Run1 - Run4} & 0.980 & 0.978 & 0.981 & 0.979 & 0.982 & 0.975 & 0.982 & 0.977  & 0.987 & 0.990\\
         
         \textbf{Run1 - Run5}  & 0.982 & 0.983 & 0.983 & 0.986 & 0.982 & 0.984 & 0.984 & 0.982 & 0.979 & 0.984\\
         
         \textbf{Run2 - Run3} & 0.991 & 0.992 & 0.991 & 0.992 & 0.991 & 0.991 & 0.991 & 0.992 & 0.997 & 0.988\\
         
         \textbf{Run2 - Run4} & 0.988 & 0.988 & 0.988 & 0.982 & 0.988 & 0.983 & 0.987 & 0.988 & 0.987 & 0.988\\
         
         \textbf{Run2 - Run5} & 0.993 & 0.993 & 0.993 & 0.991 & 0.993 & 0.994 & 0.994 & 0.993  & 0.977 & 0.981\\
         
         \textbf{Run3 - Run4} & 0.988 & 0.987 & 0.987 & 0.983 & 0.989 & 0.986 & 0.986 & 0.987 & 0.993 & 0.988\\
         
         \textbf{Run3 - Run5} & 0.992 & 0.990 & 0.993 & 0.993 & 0.993 & 0.993 & 0.992 & 0.993 & 0.981 & 0.978\\
         
         \textbf{Run4 - Run5} & 0.991 & 0.991 & 0.991 & 0.989 & 0.991 & 0.988 & 0.993 & 0.992  & 0.982 & 0.977\\
         \midrule
         \textbf{Average} & \textbf{0.988} & \textbf{0.988}& \textbf{0.988}& \textbf{0.987}& \textbf{0.988}& \textbf{0.986}& \textbf{0.989}& \textbf{0.988} & \textbf{0.987}& \textbf{0.984}\\
         \midrule
    \end{tabular}
    }
    \label{tab:ReproTablePerCondition}
\end{table*}

\begin{table*}[htb]
    \centering
    \caption{Pearson correlation between different runs of the reproducibility test in clip level.} 
    \resizebox{\columnwidth}{!}{%
    \begin{tabular}{c |c c c c c c c c | c c}
         \toprule
         & \multicolumn{8}{c}{\textbf{Template A}} &\multicolumn{2}{|c}{\textbf{Template B}} \\
         & & \textbf{Comfortable} & \textbf{Comfortable} & & & & & &\textbf{Emotion} &\\
          & \textbf{Appropriate} & \textbf{Interacting} & \textbf{Using} & \textbf{Formal} & \textbf{Affinity} & \textbf{Not Creepy} & \textbf{Realistic} & \textbf{Trust} & \textbf{accuracy} &\textbf{Resemblance}\\          
         \midrule         
         \textbf{Run1 - Run2} & 0.952 & 0.949 & 0.952 & 0.960& 0.950 & 0.940 & 0.953 & 0.950 & 0.849 & 0.962\\
         
         \textbf{Run1 - Run3} & 0.966 & 0.964 & 0.968 & 0.971 & 0.966 & 0.951 & 0.969 & 0.967 & 0.863 & 0.963 \\
         
         \textbf{Run1 - Run4} & 0.941 & 0.937 & 0.946 & 0.940 & 0.942 & 0.930 & 0.946 & 0.942 &  0.828 & 0.954\\
         
         \textbf{Run1 - Run5} & 0.954 & 0.954 & 0.955 & 0.959 & 0.954 & 0.952 & 0.957 & 0.954 & 0.835 & 0.960\\
         
         \textbf{Run2 - Run3} & 0.974 & 0.976 & 0.974 & 0.976 & 0.974 & 0.973 & 0.976 & 0.975 &  0.921 & 0.963\\
         
         \textbf{Run2 - Run4} & 0.957 & 0.956 & 0.955 & 0.948 & 0.956 & 0.951 & 0.959 & 0.955 & 0.892 & 0.953\\
         
         \textbf{Run2 - Run5} &0.973 & 0.973 & 0.971 & 0.969 & 0.972 & 0.972 & 0.974 & 0.971 &  0.849 & 0.957\\
         
         \textbf{Run3 - Run4} & 0.958 & 0.956 & 0.958 & 0.952 & 0.958 & 0.957 & 0.957 & 0.957 & 0.919 & 0.960\\
         
         \textbf{Run3 - Run5}  & 0.980 & 0.978 & 0.981 & 0.980 & 0.982 & 0.982 & 0.979 & 0.979 & 0.875 & 0.961\\
         
         \textbf{Run4 - Run5} & 0.966 & 0.965 & 0.964 & 0.966 & 0.967 & 0.966 & 0.970 & 0.966 & 0.870 & 0.951\\
         \midrule
         \textbf{Average} &\textbf{ 0.962}& \textbf{0.961}& \textbf{0.962}& \textbf{0.962}& \textbf{0.962}& \textbf{0.957}& \textbf{0.964} & \textbf{0.962} & \textbf{ 0.870} & \textbf{0.958}\\
         \midrule
    \end{tabular}
    }
    \label{tab:ReproTablePerClip}
\end{table*}

\subsection{Results}

\begin{figure*}
    \centering
    \subfloat[]{
  \includegraphics[width=\columnwidth]{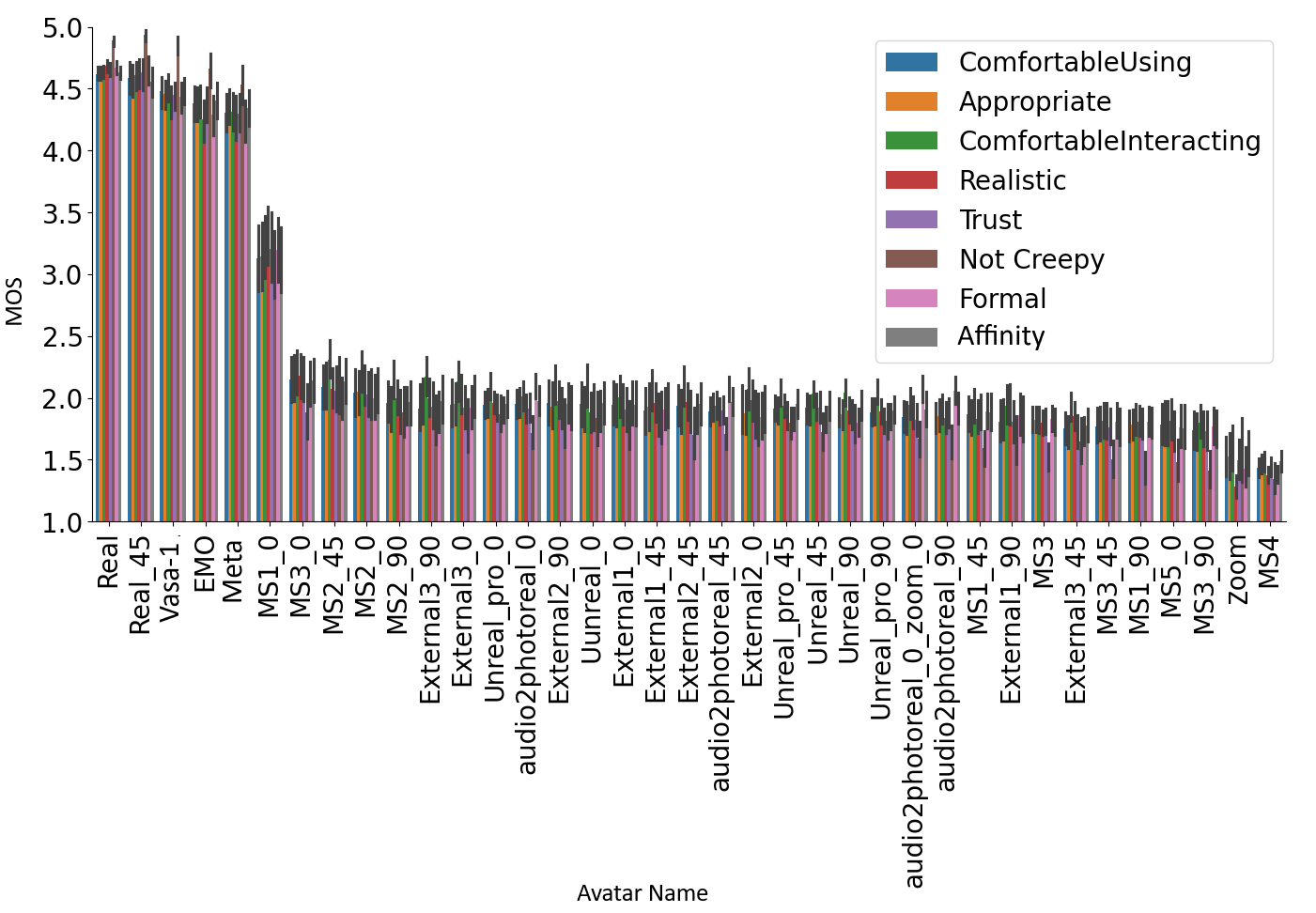}  
  }
  
  \subfloat[]{ \includegraphics[width=\columnwidth]{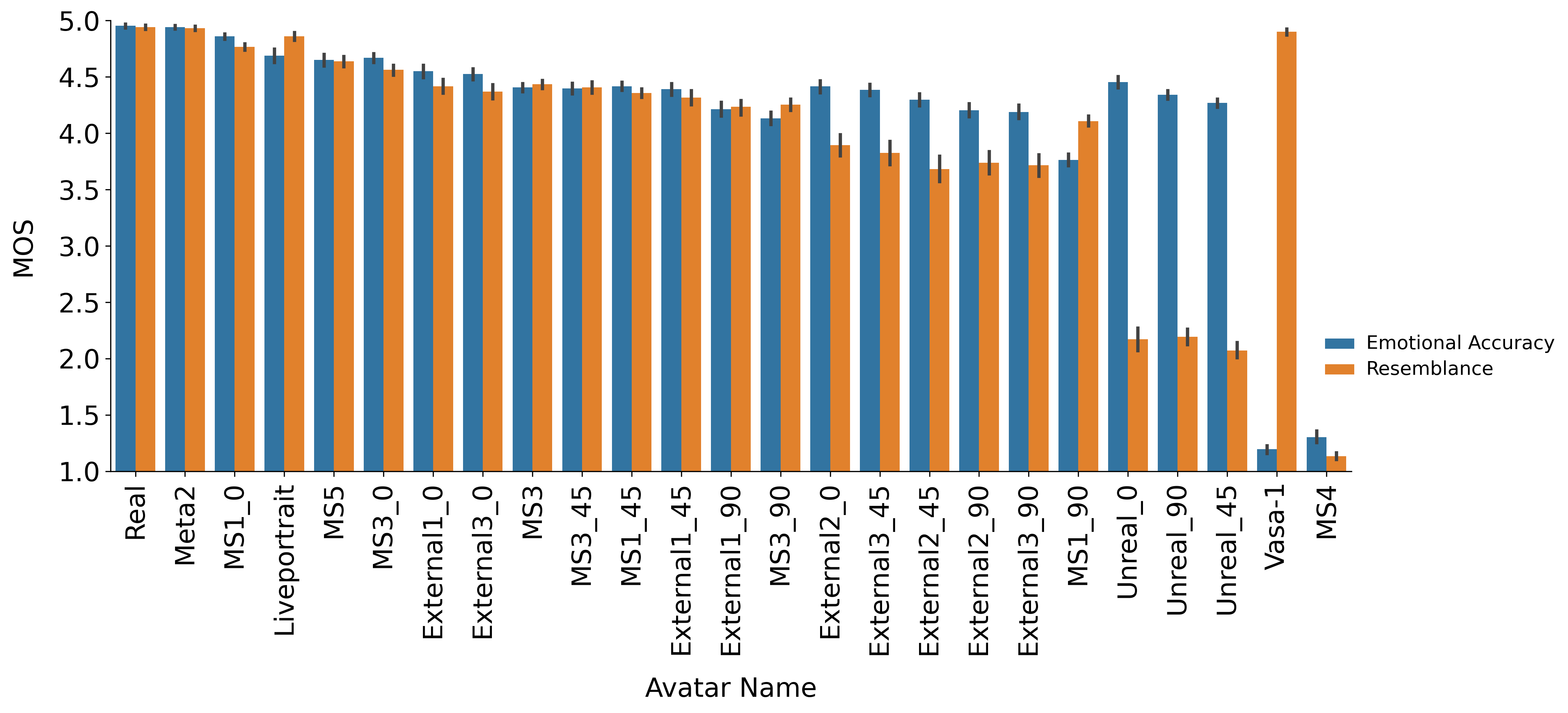}
  }
  
  \caption{(a) MOS scores per dimension across all models from Template A and (b) Template B.}
  \label{fig:MOS_Scores_Model_AllAngles}
\end{figure*}

\begin{figure*}
    \centering
  \includegraphics[width=\columnwidth]{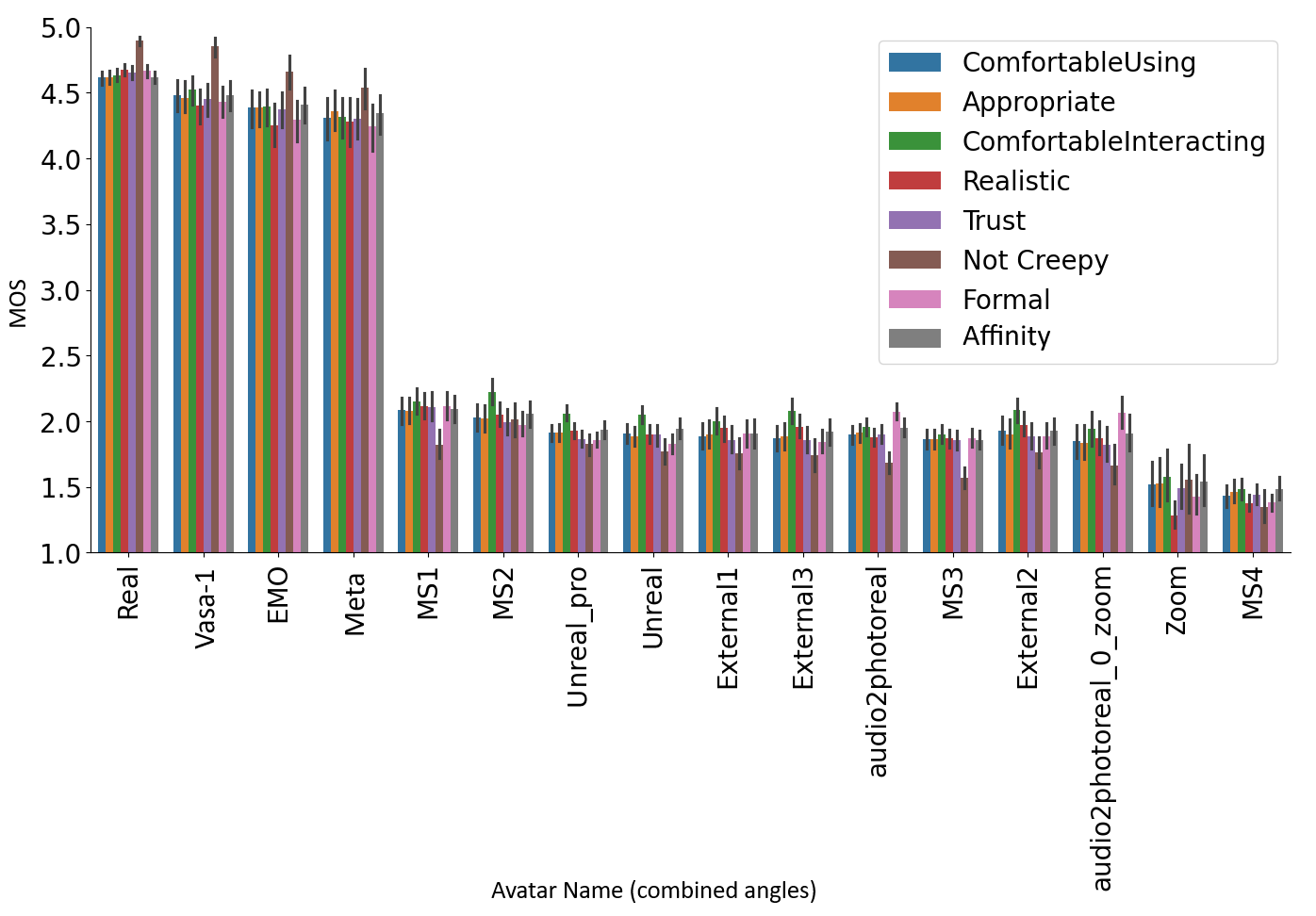}
  \caption{Template A MOS scores per dimension across all models irrespective of the view angle.}
  \label{fig:MOS_Scores_Model_NoAngles}
\end{figure*}

\begin{figure*}
    \centering
    \includegraphics[width=\columnwidth]{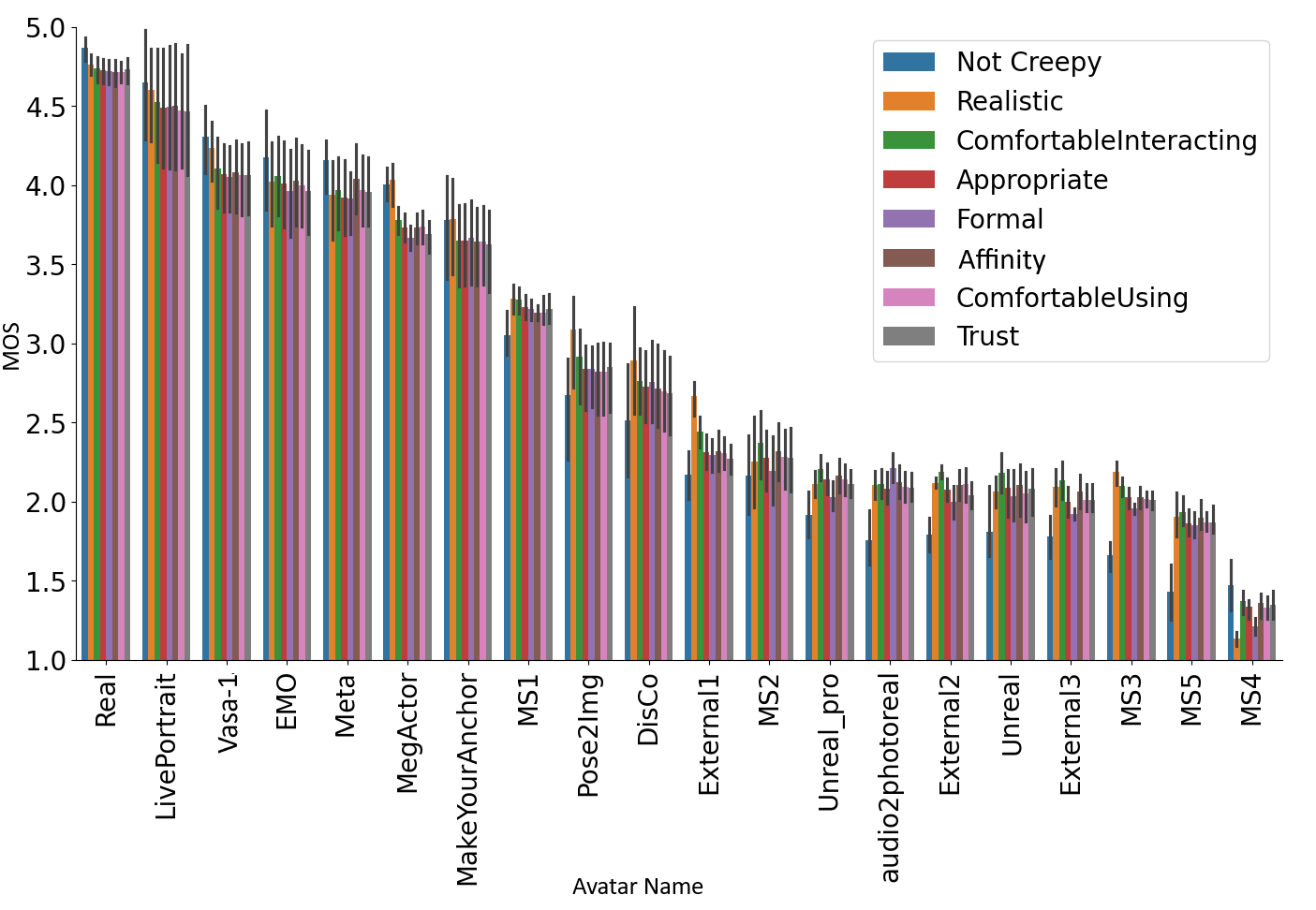}
    \caption{MOS scores per dimension across all models - No audio in clips}
    \label{fig:MOS_Scores_Model_NoAudio}
\end{figure*}


\begin{table*}[htb]
    \centering
    \caption{\markit{PCC for dimensions from Template A and B at model level.}} 
    \resizebox{\columnwidth}{!}{%
    \begin{tabular}{l | c c c c c c c c | c c}
         \toprule
         & \multicolumn{8}{c}{\textbf{Template A}} &\multicolumn{2}{|c}{\textbf{Template B}} \\
         & & \textbf{Comfortable} & \textbf{Comfortable} & & & & & &\textbf{Emotion} &\\
          & \textbf{Appropriate} & \textbf{interacting} & \textbf{using} & \textbf{Formal} & \textbf{Affinity} & \textbf{Not Creepy} & \textbf{Realistic} & \textbf{Trust} & \textbf{accuracy} &\textbf{Resemblance}\\          
         \midrule         
         \textbf{Appropriate} & 1.000 & 0.997 & 1.000 & 0.998 & 1.000 & 0.999 & 0.997 & 1.000 & 0.650 & 0.390\\
         
         \textbf{Comfortable interacting} & 0.997 & 1.000 & 0.995 & 0.993 & 0.996 & 0.997 & 0.996 & 0.996 & 0.651 & 0.407 \\
         
         \textbf{Comfortable using} & 1.000 & 0.995 & 1.000 & 0.998 & 0.999 & 0.998 & 0.997 & 1.000 &  0.653 & 0.398\\
                  
         \textbf{Formal} & 0.998 & 0.993 & 0.998 & 1.000 & 0.997 & 0.999 & 0.997 & 0.998 & 0.659 & 0.406\\
         
         \textbf{Affinity} & 1.000 & 0.996 & 0.999 & 0.997 & 1.000 & 0.998 & 0.995 & 0.999 &  0.636 & 0.372\\
         
         \textbf{Not Creepy} & 0.999 & 0.997 & 0.998 & 0.999 & 0.998 & 1.000 & 0.997 & 0.998 & 0.646 & 0.385\\
         
         \textbf{Realistic} &0.997 & 0.996 & 0.997 & 0.997 & 0.995 & 0.997 & 1.000 & 0.997 &  0.669 & 0.454\\

         \textbf{Trust} & 1.000 & 0.996 & 1.000 & 0.998 & 0.999 & 0.998 & 0.997 & 1.000 & 0.656 & 0.404\\
         \midrule
         \textbf{Emotion accuracy}  & 0.650 & 0.651 & 0.653 & 0.659 & 0.636 & 0.646 & 0.669 & 0.656 & 1.000 & 0.596\\
         
         \textbf{Resemblance} & 0.390 & 0.407 & 0.398 & 0.406 & 0.372 & 0.385 & 0.454 & 0.404 & 0.596 & 1.000\\         
     
         \midrule
    \end{tabular}
    }
    \label{tab:PerModelCorrelation}
\end{table*}

After all of the validation checks we plot the results of the survey in three formats shown in Figures \ref{fig:MOS_Scores_Model_AllAngles},  \ref{fig:MOS_Scores_Model_NoAngles}, and \ref{fig:MOS_Scores_Model_NoAudio}. Figure \ref{fig:MOS_Scores_Model_NoAngles} combines the results across various viewing angles for each avatar into a single average score per item, whereas Figure \ref{fig:MOS_Scores_Model_AllAngles} shows the average scores for each of the viewing angles. Because some models such as Make-Your-Anchor \cite{huang_make-your-anchor_2024} do not include audio, we include Figure \ref{fig:MOS_Scores_Model_NoAudio} which shows the average scores for all models without audio. For clarity, we complement the score of ``Creepy'' to show it as ``Not Creepy'' in all of the plots and tables. The avatar names ending with ``\textunderscore90''
 or ``\textunderscore45'' denote the avatars at \(90^\circ\) and \(45^\circ\) viewing angle respectively. The avatar names ending with \textunderscore0 denote frontal view. The ``audio2photoreal\textunderscore0\textunderscore zoom'' avatar is a zoomed in version of the ``audio2photoreal'' avatar clip that masks the deformations in the hands and torso.

\subsection{Correlation analysis for Template A}
We observed a strong correlation between the dimensions from Template A (Table~\ref{tab:PerModelCorrelation}), on average PCC=0.996. This was surprisingly high, which motivated doing the three versions of the survey in Section \ref{sec:survey_comparisons} which all exhibited the same high correlation. However, when we consider only low realism avatars (Realistic $\leq$ 2) then the dimensions are not all highly correlated as shown in Table \ref{tab:CorrelationAcrossDimensionsLt2}. But for avatars with Realistic > 2, the dimensions are very highly correlated with an average PCC=0.999 as shown in Table \ref{tab:CorrelationAcrossDimensionsGt2}. One important implication of this high correlation is that  avatars that are not as realistic as real video will have lower trust, comfortableness using, comfortableness interacting with, appropriateness for work, formality, and affinity, and higher creepiness compared to real video.

Figure \ref{fig:sd_realism} shows the standard deviation across all dimensions in Template A versus realism. This shows that the standard deviation significantly decreases as the avatars become more realistic.

\begin{table*}[htb]
    \centering
    \caption{\markit{PCC across dimensions when realism > 2.}} 
    \resizebox{\columnwidth}{!}{%
    \begin{tabular}{l  c c c c c c c c }
         \toprule
         & & \textbf{Comfortable} & \textbf{Comfortable} & & & & & \\
          & \textbf{Appropriate} & \textbf{interacting} & \textbf{using} & \textbf{Formal} & \textbf{Affinity} & \textbf{Not Creepy} & \textbf{Realistic} & \textbf{Trust} \\          
         \midrule
         \textbf{Appropriate} & 1.000 & 0.998 & 0.999 & 0.997 & 0.999 & 0.996 & 0.997 & 0.999 \\
         
         \textbf{Comfortable interacting} & 0.998 & 1.000 & 0.998 & 0.996 & 0.998 & 0.998 & 0.997 & 0.997  \\
         
         \textbf{Comfortable using} & 0.999 & 0.998 & 1.000 & 0.997 & 0.998 & 0.996 & 0.996 & 0.998 \\
                  
         \textbf{Formal} & 0.997 & 0.996 & 0.997 & 1.000 & 0.997 & 0.993 & 0.997 & 0.998 \\
         
         \textbf{Affinity} & 0.999 & 0.998 & 0.998 & 0.997 & 1.000 & 0.996 & 0.996 & 0.998\\
         
         \textbf{Not Creepy} & 0.996 & 0.998 & 0.996 & 0.993 & 0.996 & 1.000 & 0.994 & 0.995\\
         
         \textbf{Realistic} &0.997 & 0.997 & 0.996 & 0.997 & 0.996 & 0.994 & 1.000 & 0.997 \\
                  
         \textbf{Trust} & 0.999 & 0.997 & 0.998 & 0.998 & 0.998 & 0.995 & 0.997 & 1.000 \\
     
         \midrule
    \end{tabular}
    }
    \label{tab:CorrelationAcrossDimensionsGt2}
\end{table*}

\begin{table*}[htb]
    \centering
    \caption{\markit{PCC across dimensions when realism $\leq$ 2}} 
    \resizebox{\columnwidth}{!}{%
    \begin{tabular}{l  c c c c c c c c }
         \toprule
         & & \textbf{Comfortable} & \textbf{Comfortable} & & & & & \\
          & \textbf{Appropriate} & \textbf{interacting} & \textbf{using} & \textbf{Formal} & \textbf{Affinity} & \textbf{Not Creepy} & \textbf{Realistic} & \textbf{Trust} \\          
         
         \midrule
         \textbf{Appropriate} & 1.000 & 0.824 & 0.909 & 0.704 & 0.885 & 0.655 & 0.763 & 0.894 \\
         
         \textbf{Comfortable interacting} & 0.824 & 1.000 & 0.854 & 0.575 & 0.879 & 0.806 & 0.771 & 0.797  \\
         
         \textbf{Comfortable using} & 0.909 & 0.854 & 1.000 & 0.693 & 0.908 & 0.681 & 0.763 & 0.879 \\
                  
         \textbf{Formal} & 0.704 & 0.575 & 0.693 & 1.000 & 0.707 & 0.384 & 0.670 & 0.706 \\
         
         \textbf{Affinity} & 0.885 & 0.879 & 0.908 & 0.707 & 1.000 & 0.732 & 0.783 & 0.871\\
         
         \textbf{Not Creepy} & 0.655 & 0.806 & 0.681 & 0.384 & 0.732 & 1.000 & 0.549 & 0.600\\
         
         \textbf{Realistic} &0.763 & 0.771 & 0.763 & 0.670 & 0.783 & 0.549 & 1.000 & 0.727 \\
                  
         \textbf{Trust} & 0.894 & 0.797 & 0.879 & 0.706 & 0.871 & 0.600 & 0.727 & 1.000 \\
     
         \hline
    \end{tabular}
    }
    \label{tab:CorrelationAcrossDimensionsLt2}
\end{table*}

\begin{figure*}
    \centering
    \includegraphics[width=\columnwidth]{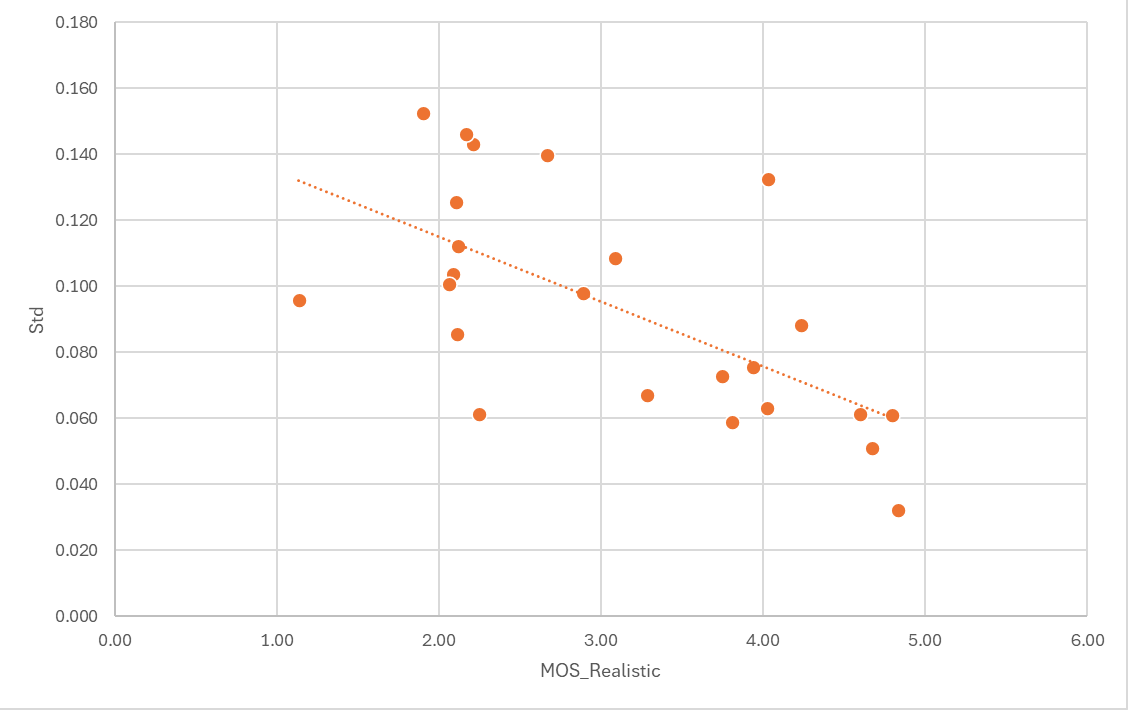}
    \caption{Standard deviation across all dimensions in Template A versus realism. A best fit line is shown with a $R^2=0.41$.}
    \label{fig:sd_realism}
\end{figure*}

\subsection{Dimensionality analysis}
We conducted Principal Component Analysis (PCA) to reduce the dimensionality of the measure items from Templates A and B. We observed two orthogonal principal components that explain 94\% of the variance in the data. Items from Template A load similarly on the first component. Items from Template B load on the second component, with Emotion Accuracy having the highest loading, and Resemblance having a cross-loading on the first component (0.462). Consequently, we adopted the two-component model. As a result, a new template with one item selected from Template A (e.g., realism) and the two items from Template B, can be used for photorealistic avatars to significantly reduce the subjective test time.

\subsection{Correlation with objective metrics}
We evaluated the effectiveness of objective metrics in representing subjective scores. In this domain, PSNR~\cite{gonzalez_digital_2006}, SSIM~\cite{wang_image_2004}, and LPIPS~\cite{zhang_unreasonable_2018} are typically reported in publications. We also analyze FID \cite{heusel_gans_2017} and FVD \cite{unterthiner_fvd_2019}, which are less commonly reported. We utilized eight models from the study reported in Section~\ref{sec:repro}, where the original video was available (overall 31 clips). We removed the background of the person from the original videos and transformed the videos to match each avatar representation using face landmarks (see Figure~\ref{fig:obj_metric_transfer_viz} for an example). Finally, we calculated the objective metrics for each video clip and aggregated them for the avatar models. Table~\ref{tab:subj_obj_head} reports the Pearson's correlation coefficient and Kendall's Tau-b rank correlation coefficient\footnote{We report Kendall's Tau-b as it is more robust to outliers and non-normal distributions compare to Spearman's rank correlation~\cite{croux2010influence} making it more suitable for small sample sizes.} \cite{kendall_new_1938} between the objective metrics and the subjective scores from Template A and Template B. The results show a weak correlation between the objective metrics and the subjective scores from Template A, and a moderate correlation with scores from Template B for PSNR, SSIM, and LPIPS. This highlights the necessity of using subjective scores for accessing the quality of experience of photorealistic avatars. 

\begin{figure*}
    \centering
    \subfloat[]{\includegraphics[width=0.33\textwidth]{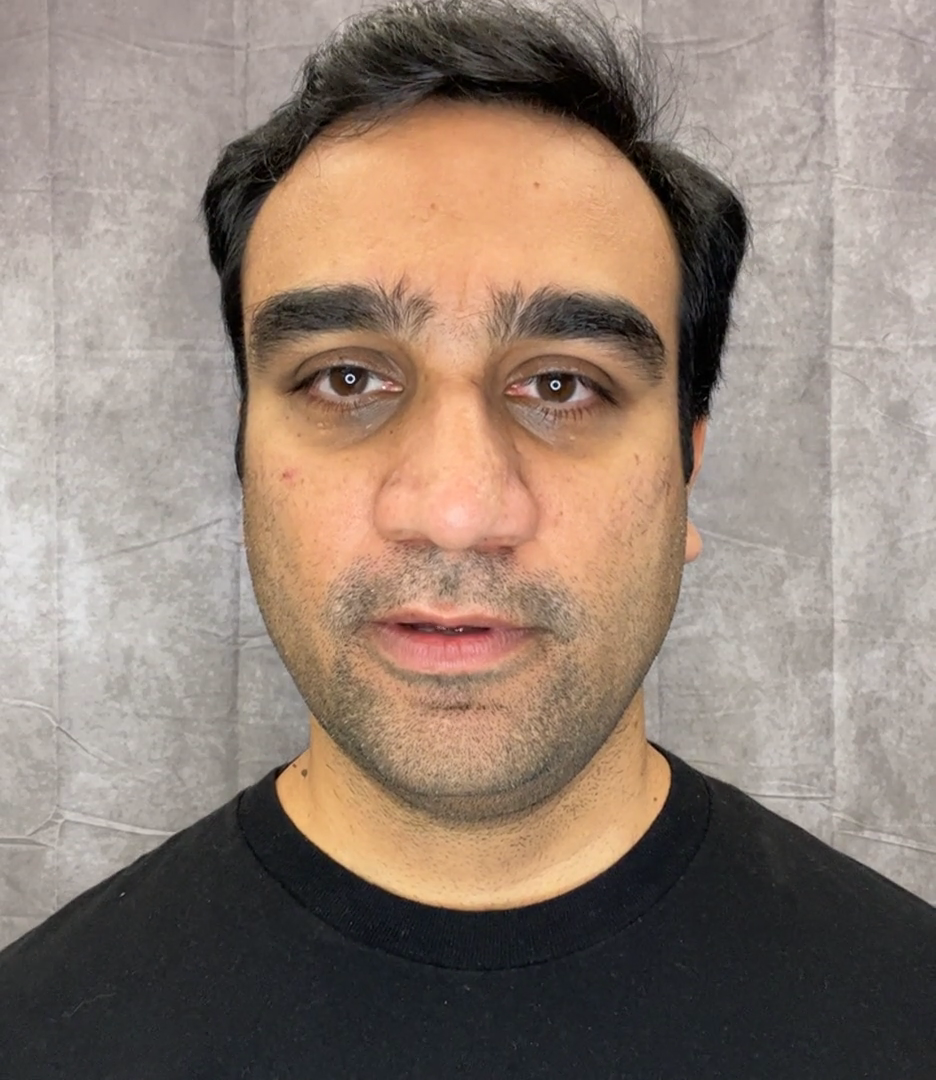}}
    \subfloat[]{\includegraphics[width=0.33\textwidth]{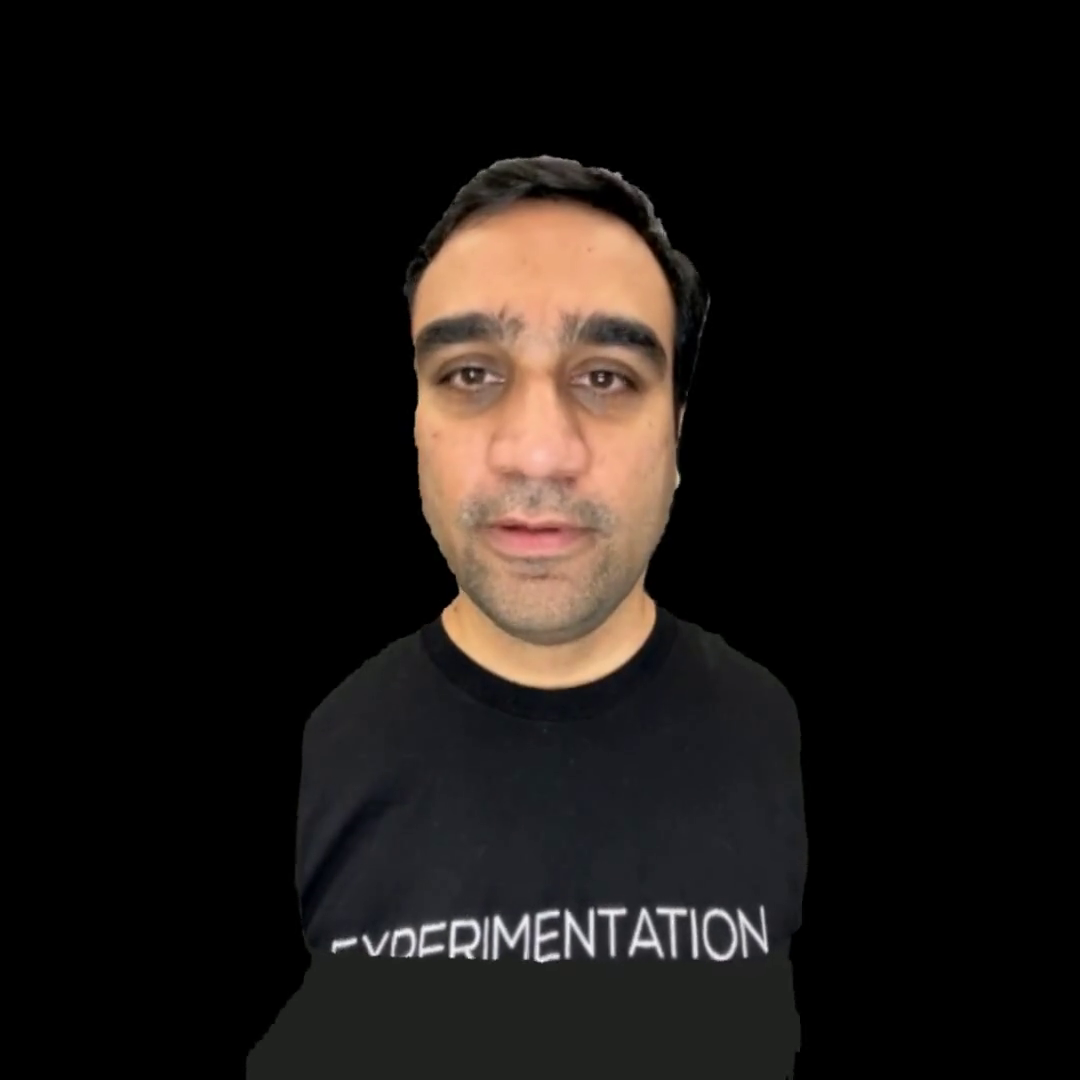}}
    \subfloat[]{\includegraphics[width=0.33\textwidth]{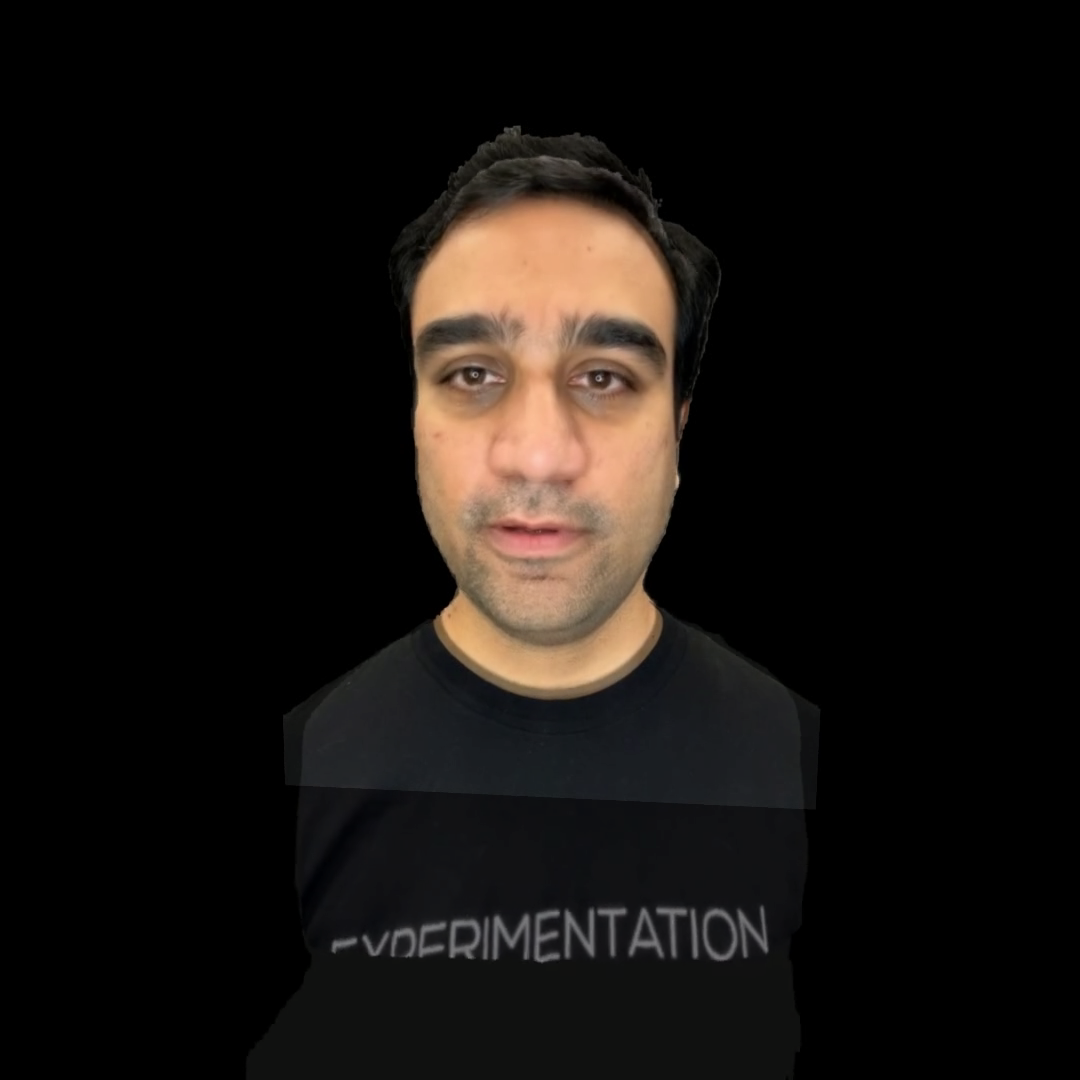}    }
    
    \caption{\markit{Modifying the original video to match the position of the avatar using face landmarks and removing the background before calculating objective metrics. (a) A frame from the original video, (b) the corresponding frame from the avatar, (c) an overlay of the modified original frame and the avatar's frame.}}
    \label{fig:obj_metric_transfer_viz}
\end{figure*}

Table~\ref{tab:subj_obj_face} shows the results for the correlation between subjective scores and objective metrics in the face only region. In this restricted case, FID provides moderate to good correlation across all dimensions. Finally, list of avatar models, the objective metrics and subjective scores are reported in Table~\ref{tab:subj_vs_objective}.  

\begin{table}[t]
\caption{Correlation between subjective scores and objective metrics calculated on \textit{head and torso} area. The highest correlations per column are marked in bold.}
\label{tab:subj_obj_head} 
\begin{center}
\resizebox{0.8\columnwidth}{!}{%
    \begin{tabular}{l c c c c c c c c c c }
    \toprule    
    & \multicolumn{5}{c}{\textbf{Pearson correlation}} & \multicolumn{5}{c}{\textbf{Kendall Tau-b correlation}} \\    
    \textbf{Items} & 
    \textbf{PSNR } & 
    \textbf{SSIM } & 
    \textbf{LPIPS} & 
    \textbf{FID} & 
    \textbf{FVD} & 
    
    \textbf{PSNR } & 
    \textbf{SSIM } & 
    \textbf{LPIPS} &
    \textbf{FID} &
    \textbf{FVD} 
    \\
    
    \midrule
    \textbf{Appropriate} & 0.130& 0.110& -0.295 & -0.408 & -0.084 & 0.286 & 0.214 & -0.286 & -0.286 & -0.286\\
    \textbf{Comfortable Interacting} & 0.094 & 0.077 & -0.262 & -0.417 & -0.078 & 0.214 & 0.143 & -0.214& \textbf{-0.357} & \textbf{-0.357}\\
    \textbf{Comfortable Using} &0.143 & 0.119 & -0.300 & -0.419 & -0.101 & 0.214 & 0.143 & -0.214 & \textbf{-0.357} & \textbf{-0.357}\\   
    \textbf{Formal} & 0.158 &  0.129 & -0.310 & -0.419 & -0.096 & 0.214 & 0.143 & -0.214 & -0.214 & -0.214\\    
    \textbf{Affinity} & 0.119 & 0.100 &  -0.287 & -0.412 & -0.091 & 0.286 & 0.214 & -0.286 & -0.286 & -0.286\\

    \textbf{Not Creepy} &0.130 &  0.105 & -0.293 & \textbf{-0.430} & -0.104 & 0.286 & 0.214 & -0.286 & -0.286& -0.286\\
    
    \textbf{Realistic} & 0.132 & 0.100 &  -0.276 & -0.426 & -0.094  & 0.214 & 0.143 & -0.214 & \textbf{-0.357}& \textbf{-0.357}\\
    
    \textbf{Trust} & 0.138 & 0.116 & -0.297 & -0.411 & -0.088 & 0.214 & 0.143 & -0.214 & \textbf{-0.357}& \textbf{-0.357}\\    
     \midrule
    \textbf{Emotion accuracy}  & \textbf{0.630} & \textbf{0.619} & \textbf{-0.670} & -0.085 & 0.067 & \textbf{0.571} &  \textbf{0.500} & \textbf{-0.429} & -0.143 & 0.000 \\    
    \textbf{Resemblance} & 0.298 & 0.182 & -0.126 & -0.345& \textbf{-0.128} & 0.286 &0.357 & \textbf{-0.429} & -0.143 & -0.143\\

    \midrule
    \end{tabular}%
}
\end{center}
\end{table}

\subsection{Uncanny valley effect}
Figure \ref{fig:AffinityVersusRealism} plots the avatar affinity versus realism for all of the avatar clips used Figure \ref{fig:MOS_Scores_Model_NoAudio}. This shows a very linear relationship between affinity and realism with an $R^2=0.966$. There is no dip in affinity that \cite{mori_uncanny_2012} predicts as the avatar realism becomes near photorealistic. This implies that photorealistic avatars should be as realistic as possible to maximize affinity. This means there is no reason to stay to the left of the uncanny value since it does not exist when not considering robot face images \cite{mathur_navigating_2016}. Because the dimensions used in Template A are all highly correlated for realism > 2 (Table \ref{tab:CorrelationAcrossDimensionsGt2}), maximizing realism also maximizes trust, comfortableness using, comfortableness interacting with, appropriateness for work, and formality, while minimizing creepiness.

\begin{figure}
    \centering
    \includegraphics[width=0.9\columnwidth]{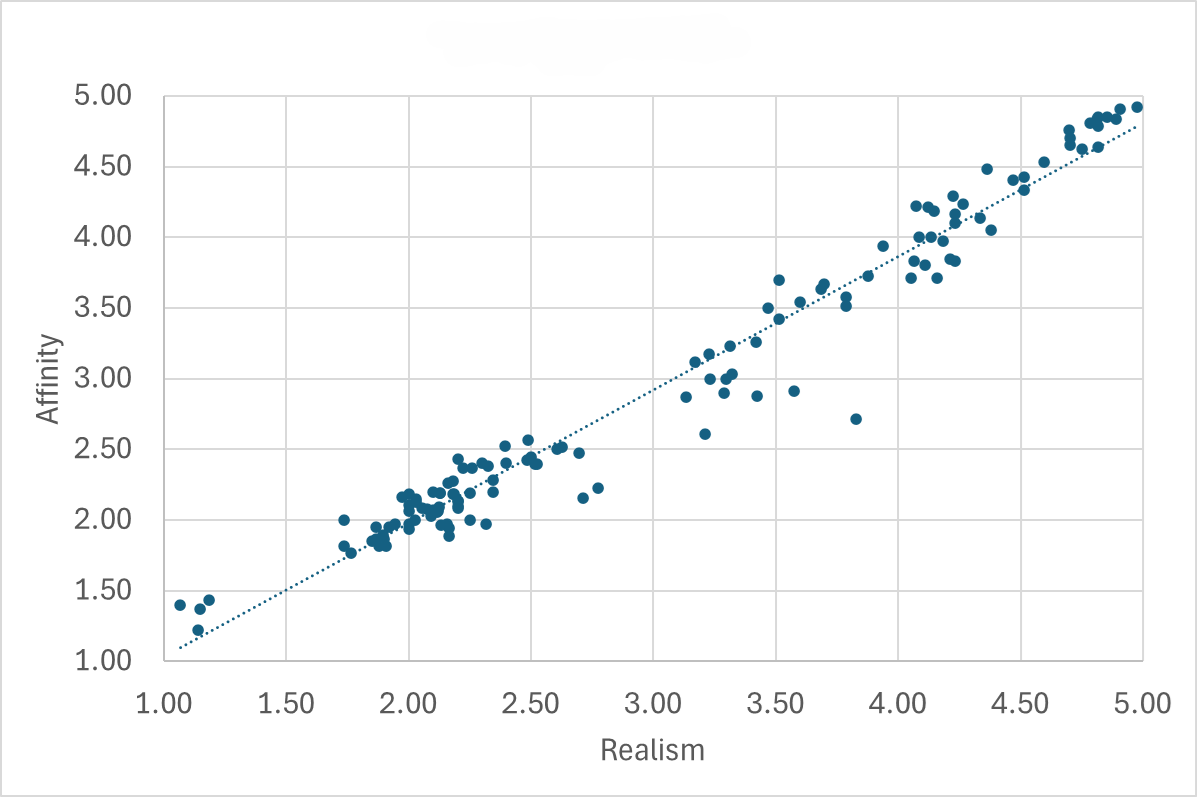}
    \caption{Affinity versus realism for the video clips in Figure \ref{fig:MOS_Scores_Model_NoAudio}. \(R^2\)=0.966}
    \label{fig:AffinityVersusRealism}
\end{figure}

\section{Conclusions and future work}
\label{sec:conclusions}
We have shown the commonly used objective metrics (PSNR, SSIM, LPIPS, FIV, FVD) are weakly correlated to nine of the subjective metrics we define, and moderately correlated to emotion accuracy (RQ1). Note that none of these metrics are trained to predict avatar realism. More sophisticated objective metrics may improve objective performance, especially if trained using a large subjective dataset of avatar realism (this is done in audio \cite{yi_conferencingspeech_2022} and video \cite{majeedi_full_2023} objective quality models). 

For avatars with a realism $>$ 2 MOS we have shown in Table \ref{tab:CorrelationAcrossDimensionsGt2} that all of the no-reference metrics (Template A) are highly correlated to each other, which means that for Template A only one metric needs to be measured (e.g., realism), while for Template B both metrics need to be measured; this leads to dimensionality reduction of 10 to 3 for the survey (RQ2).

We have provided an open source crowdsourced test framework to measure the quality of experience of photorealistic avatars, and shown it to be accurate and reproducible (RQ3).

We have shown in the context of photorealistic avatars in telecommunication scenarios that the uncanny valley effect does not exist, which implies photorealistic avatars should be made as realistic as possible to maximize the quality of experience of the avatar (RQ4). Note that uncanny valley effects may still exist in other domains (e.g., gaming, metaverse) and with other visual distortions (e.g., motion artifacts or mismatches to expected expressions).    

Finally, we show that avatars that are not as realistic as real video will have lower trust, comfortableness using, comfortableness interacting with, appropriateness for work, formality, and affinity, and higher creepiness compared to real video (RQ5). One implication is that to get the level trust, comfortableness using, comfortableness interacting with, appropriateness for work, formality, and affinity, and higher creepiness as real video, the avatar needs the same level of realism as real video, which is a very high bar not yet achieved by any of the avatars tested. 

\subsection{Extensions}
There are many potential extensions of this work. For example: 

\begin{itemize}
    \item For avatars that show the upper half of the body (e.g., \cite{huang_make-your-anchor_2024}) we can include a question in Template B for gesture accuracy. 
    \item To better understand what areas to improve for realism, we can ask the rater to fill in why they gave it that rating and provide binary options such as (1) The avatar was distorted, (2) The lip sync was off, (3) Absence of microdetails and imperfections, (4) Inaccurate lighting and shadows, (5) Unnatural textures and materials, and (6) Other (the rater provides a description).
    \item With much more data collected using this test framework we could develop objective metrics that are highly correlated to these subjective metrics, which would further increase the development speed of photorealistic avatars. In particular, the strong correlation of the Template A metrics means we only need to develop an objective metric for say realism and not trust or appropriate for work, which seems much more tractable. 
    \item Photorealistic avatars must have the same lipreading performance as real video, not just for accessibility but to improve the intelligibility for users of the avatar. 
    \item The ratings of people we are familiar with and total strangers will likely be different, with the former being more challenging. Similarly, a self-assessment test will also likely be more challenging than raters judging people they do not know. 
    \item This test framework is a passive test, meaning the raters are not interacting with the avatars themselves, but only as an observer. A more challenging test is to let raters interact with a real-time avatar driven by a real person.
    \item The test framework could be extended to other scenarios mentioned in Section \ref{sec:introduction}, such as health care, education, retail and e-commerce, and entertainment. This would include expanding the test set of avatars, and potentially adding some new test dimensions.
    \item The test framework could be extended to 3D environments such as virtual reality (VR) headsets. 
    \item The uncanny valley effect can be further studied using attractiveness, eeriness, humanness, and spine-tingling.
\end{itemize}

\subsection{Developer guidelines}
This study provides some specific guidelines for developers of telecommunication systems that use photorealistic avatars:

\begin{itemize}
    \item The avatars should be as realistic as possible, ideally matching the realism from real video. This will maximize trust, comfortableness using, comfortableness interacting with, appropriateness for work, formality, and affinity, while minimizing creepiness. Critically, avatars that are not as realistic as real video will have lower trust, comfortableness using, comfortableness interacting with, appropriateness for work, formality, and affinity, and higher creepiness compared to real video.
    \item The avatars should maximize resemblance to the person, ideally being indistinguishable from real video. This will maximize the identity recognition of the person.
    \item The avatars should maximize emotion accuracy of the person, ideally being indistinguishable from real video. This will maximize the communication content transmission for the collaboration system. 
    \item During the development of collaboration systems with photorealistic avatars regular subjective tests should be conducted as we have shown the objective metrics are not well correlated to the subjective metrics in this study. Not all ten dimensions need to be measured as we have shown only 3 are needed. This greatly simplifies the measurement process as well as reducing the number of dimensions that need to be optimized during development.  
\end{itemize}

\subsection{Limitations}
There are several limitations for this study, including:

\begin{itemize}
    \item The number of avatars used in the study (N=19) would ideally be much larger (e.g., N=50). In addition, the avatars should ideally be all tested with identical material, which was only the case when the avatar model was available. 
    \item The diversity of the avatars studied: We used 19 avatar models to create 88 instances of avatars, which included 67 males, 21 females, 51 Caucasians, 27 Asians, and 10 Blacks. Ideally, we would have included more avatars with an equal number of males and females and a greater number of and more equally distributed races represented. 
    \item The test framework is a passive test and the results of actively using the avatars may be different.
    \item The evaluations are done on 2D displays and the results may be different if tested on 3D displays or VR headsets.
    \item The use of photorealistic avatars in telecommunication systems introduces potential privacy, trust, and ethical considerations, which are outside the scope of this paper and should be addressed by the community. 
\end{itemize}

\begin{acks}
We thank various teams at Microsoft for providing some avatars for this study.
\end{acks}

\bibliographystyle{ACM-Reference-Format}
\bibliography{IC3-AI,other}
\appendix 
\section{Appendix}
Table~\ref{tab:subj_obj_face} presents the correlation between subjective and full-reference objective metrics when calculated only in the face area.

\begin{table}[htb]
\caption{Correlation between subjective scores and objective metrics only \textit{on face area}. The highest correlations per column are marked in bold.}
\label{tab:subj_obj_face}
\begin{center}
\resizebox{0.8\columnwidth}{!}{%
    \begin{tabular}{l c c c c c c c c c c}
    \toprule
    & \multicolumn{5}{c}{\textbf{Pearson correlation}} & \multicolumn{5}{c}{\textbf{Kendall Tau-b correlation}} \\
    \textbf{Items} & 
    \textbf{PSNR } & 
    \textbf{SSIM } & 
    \textbf{LPIPS} & 
    \textbf{FID} & 
    \textbf{FVD} & 
    \textbf{PSNR } & 
    \textbf{SSIM } & 
    \textbf{LPIPS} &
    \textbf{FID} & 
    \textbf{FVD} \\

    \midrule
    \textbf{Appropriate} & 0.287 & 0.323 & -0.477 & -0.765 & -0.101 & 0.214 & 0.143 & -0.214 & -0.286 & 0.000 \\
    \textbf{Comfortable Interacting} & 0.274 & 0.296 & -0.476 & \textbf{-0.796} & -0.101 & 0.286 & 0.214 & -0.286 & -0.357 & -0.071 \\ 
    \textbf{Comfortable Using} &  0.304 & 0.344 & -0.488 & -0.761 & -0.113 & 0.286 & 0.214 & -0.286 & -0.357 & -0.071 \\
    \textbf{Formal} & 0.308 & 0.351 & -0.498 & -0.748 & -0.142 & 0.143 & 0.214 & -0.143 & -0.214 & -0.071 \\
    \textbf{Affinity} & 0.273 & 0.310 & -0.460 & -0.757 & -0.082 & 0.214 & 0.143 & -0.214 & -0.286 & 0.000 \\
    \textbf{Not Creepy} &  0.285 & 0.318 & -0.476 & -0.754 & -0.110 & 0.214 & 0.143 & -0.214 & -0.286 & 0.000 \\
    \textbf{Realistic} & 0.319 & 0.364 & -0.519 & -0.788 & -0.173 & 0.286 & 0.214 & -0.286 & -0.357 & -0.071 \\
    \textbf{Trust} & 0.300 & 0.339 & -0.490 & -0.770 & -0.116 & 0.286 & 0.214 & -0.286 & -0.357 & -0.071 \\
     \midrule
    \textbf{Emotion accuracy}  & \textbf{0.764} & 0.564 & \textbf{-0.877} & -0.687 & -0.420 & \textbf{0.643} & 0.429 & \textbf{-0.786} & -0.571 & \textbf{-0.286} \\
    \textbf{Resemblance} & 0.677 & \textbf{0.781} & -0.830 & -0.688 & \textbf{-0.856} & 0.500 & \textbf{0.571} & -0.643 & \textbf{-0.714} & \textbf{-0.286} \\
    \midrule

    \end{tabular}%
}
\end{center}
\end{table}

\begin{table*}[htb]
    \centering
    \caption{Subjective scores as measured by our survey and corresponding objective scores.} 
    \includegraphics[width=\linewidth]{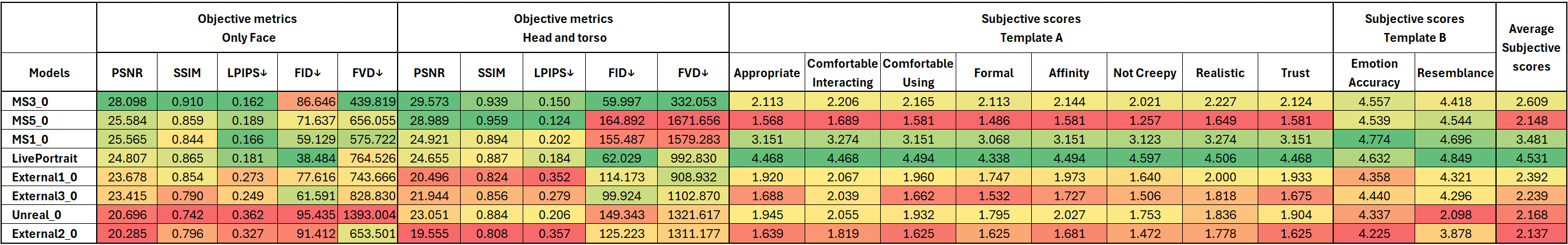}
     \label{tab:subj_vs_objective}
\end{table*}

\end{document}